\documentclass[aps,physrev,twocolumn,groupedaddress]{revtex4-2}

\usepackage{amsmath}
\usepackage{amssymb}
\usepackage{graphicx}
\usepackage{bbold}
\usepackage{hyperref}
\usepackage{algorithm}
\usepackage{algpseudocodex}
\newcommand{\ket}[1]{|#1\rangle}
\newcommand{\bra}[1]{\langle#1|}

\newcommand{\vev}[1]{\langle#1\rangle}
\renewcommand{\a}{\hat{a}}
\newcommand{\ai}{\hat{a}_i}
\newcommand{\aidag}{\hat{a}_i^\dagger}
\renewcommand{\b}{\hat{b}}
\newcommand{\bi}{\hat{b}_i}
\newcommand{\bidag}{\hat{b}_i^\dagger}
\renewcommand{\v}{\hat{v}}
\newcommand{\vi}{\hat{v}_i}
\newcommand{\vidag}{\hat{v}_i^\dagger}

\begin{document}


\title{Symmetric preferences, asymmetric outcomes: Tipping dynamics in an open-city segregation model}


\author{Fabio van Dissel}\email{fvdissel@ifae.es}
\affiliation{Institut de F\'{\i}sica d'Altes Energies (IFAE)\\ 
 The Barcelona Institute of  Science and Technology (BIST)\\
 Campus UAB, 08193 Bellaterra (Barcelona) Spain}
\author{Tuan Minh Pham}\email{m.t.pham@uva.nl}
\affiliation{Institute for Theoretical Physics, University of Amsterdam}
\affiliation{Dutch Institute for Emergent Phenomena, University of Amsterdam}
\affiliation{Complexity Science Hub Vienna, Metternichgasse 8, A-1030, Vienna, Austria}
\author{Wout Merbis}\email{w.merbis@uva.nl}
\affiliation{Dutch Institute for Emergent Phenomena, Netherlands}
\affiliation{Institute for Theoretical Physics, University of Amsterdam}
\affiliation{Department of Mathematics, VU Amsterdam}
\affiliation{Centraal Bureau voor de Statistiek, Henri Faasdreef 312, The Hague}


\date{\today}

\begin{abstract}
Schelling's model of segregation demonstrates that even in the absence of social or governmental interventions, individuals with \emph{mild} in-group preferences can self-organize into strongly segregated neighborhoods.
Many variants of this celebrated model  have been proposed by assuming agents tend to increase their satisfaction. 
Complementary to this traditional, utility-based approach, we model residential moves using \emph{satisfaction-independent} reaction rates in a spatially extended chemical reaction network.
The resulting model exhibits an emergent phenomenon: despite symmetric in-group preferences, the system undergoes a tipping transition at a critical preference level, beyond which one agent type dominates. 
We characterize this asymmetric phase transition in details using mean-field analysis, numerical simulations and finite size scaling methods. 
We find that while the transition shares key features with the Ising universality class, such as $\mathbb{Z}_2$ symmetry breaking and similar exponent ratios, the full set of critical exponents does not match any known universality class.
\end{abstract}


\makeatletter
\let\saved@author\@author
\let\saved@affiliation\@affiliation
\makeatother

\maketitle

\section{\label{sec:intro}Introduction}

Segregation is a widespread urban phenomenon in which a population is divided into subgroups along socioeconomic lines, such as ethnicity, income, education, or social economic status \cite{Schelling1969}. Since the pioneering work of Thomas Schelling \cite{schelling1971dynamic}, numerous studies have confirmed the robustness of segregation patterns arising from individual, satisfaction-driven movements \cite{Pancs, Clark, Clark_Fossett}. While early works rely on agent-based simulations, offering a mechanistic view but limited theoretical insight, recent research has revealed intriguing analogies between the Schelling model and various physical phenomena, such as surface tension \cite{Vinkovic19261} and clustering in Ising- or Blume–Emery–Griffiths-type systems \cite{stauffer2007ising,dall2008statistical,gauvin2009phase, gauvin2010schelling}. 
Such behavior is consistent with the broader class of nonequilibrium systems exhibiting Ising-type criticality, despite the absence of detailed balance \cite{odor2004universality}.
Moreover, Zakine et.~al.~\cite{zakine2024socioeconomic}  suggest that the large-scale organization seen in social segregation may follow universal scaling laws similar to those observed in active-matter systems. Seara et.~al.~\cite{seara2023sociohydrodynamics} propose a hydrodynamic model with utility-driven diffusion whose parameters are calibrated by the US census data.

In essence, these approaches rely on modeling, in one way or another, agents' preferences for  neighborhoods of a certain demographic composition. Such preferences are often described by  a so-called tolerance threshold -- the maximal fraction of neighbors of another type which agents can tolerate [or transversely: an acceptance threshold defined as the minimal fraction of like neighbors below which agents consider moving out]. The key conclusion of Schelling-type models is that \emph{mild} in-group preference for same-type neighbors is sufficient to lead to segregation. Further results confirm   this paradoxical behavior, establishing that a higher degree of tolerance can actually lead to an overall increase in segregation  \cite{Durrett,Immorlica, Barret}. Even if agents have a  preference for perfect integration, a best-response  dynamics (where  agents move if and only if they strictly improve their utility), may still lead to segregation \cite{Pancs}. In this regard, a central open question is whether segregation can arise in the \emph{absence} of such explicit thresholds or utility-optimization rules.

\begin{figure}
    \centering
    \includegraphics[width=\linewidth]{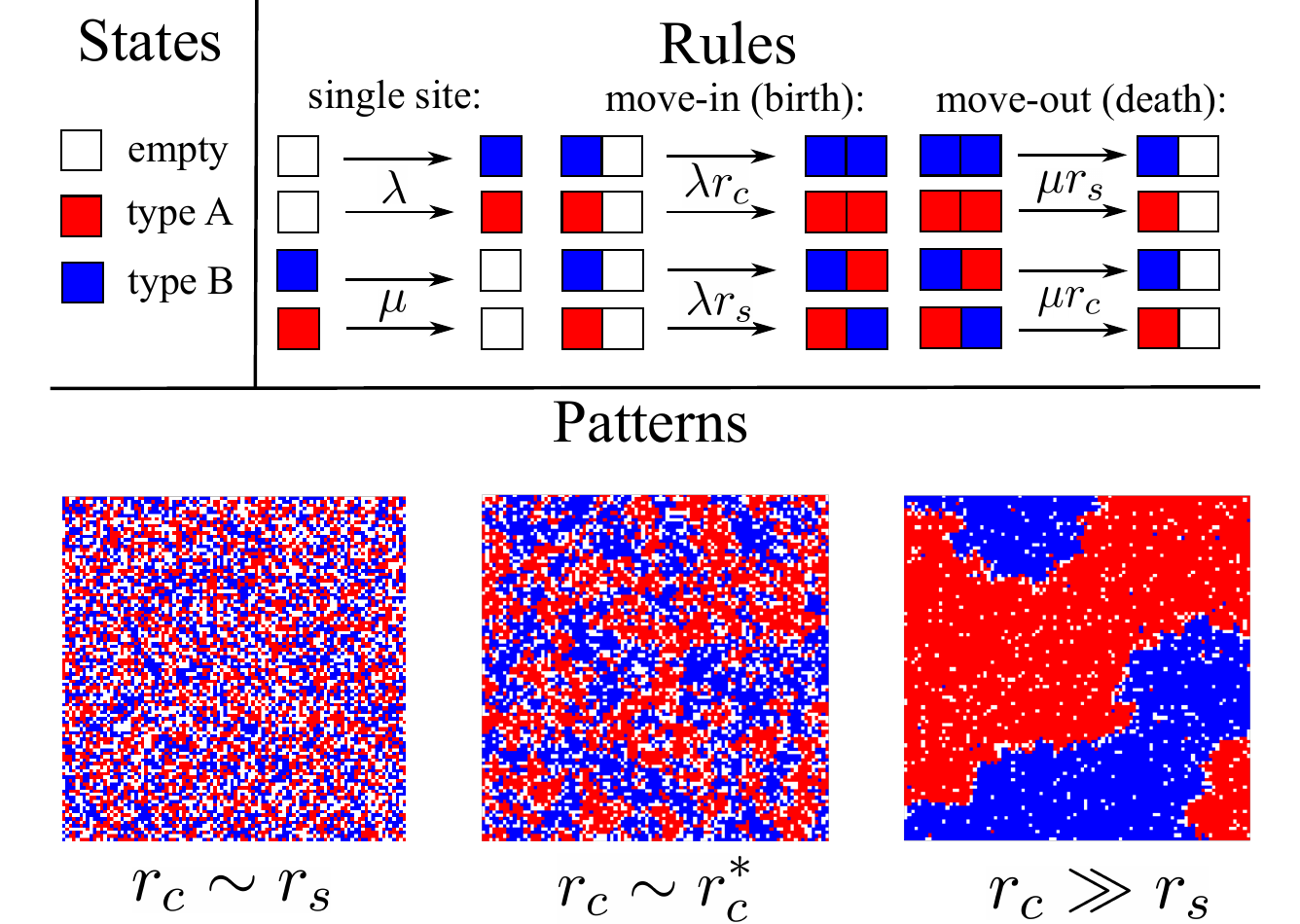}
    \caption{Overview of the reactions of our segregation model. Each lattice site can be in three possible states, corresponding to either an empty house or a resident of type $A$ or $B$. The reaction rules allow for random creation and annihilation of residents (moving into the region with rate $\lambda$ or moving out with rate $\mu$), alongside pairwise reactions grouped into move-in (birth) and move-out (death) reactions. For the pairwise moving-in reactions, we distinguish those which promote like neighbors (in-group, or `copy' with rate $\lambda r_c$) from those that promote opposing neighbors (out-group, or `split' with rate $\lambda r_s$). Similarly, we divide the pairwise moving-out reactions into two types: in-group reactions with rate $\mu r_s$ and out-group reactions with rate $\mu r_c$. Some exemplary grid configurations are shown on a 2D lattice with $100\times 100$ sites. Increasing the in-group reaction rate $r_c$ gradually leads to a segregating transition, where regions of both red and blue types are equally large. Further increasing $r_c$ will lead to larger regions of one type dominating over smaller islands of the opposing type.}
    \label{fig:overview}
\end{figure}

This situation is often met in practice. Indeed, individual motivations and preferences are typically latent and difficult to infer, whereas aggregate quantities such as residential mobility rates are often accessible through administrative records and statistical data. Motivated by the observation that such frequencies can be interpreted as reaction rates in a spatially-extended chemical reaction network (CRN),  we model the hopping dynamics using two pairwise reaction processes: in-group preference and out-group tolerance. The former encodes the tendency of agents to move into neighborhoods populated by others of the same type, while the latter captures the acceptance of agents of the opposite type into those neighborhoods. Each of these processes is accompanied by a \emph{reverse} move, where one of the two interacting agents leaves the neighborhood (see Fig. \ref{fig:overview}). Within this picture, segregation emerges as a consequence of some random reactions occurring more (or less) frequently than others, generating a systematic tendency for same-type agents to cluster together. This mechanism is closely related to domain coarsening and phase-ordering processes studied in statistical physics \cite{bray2002theory}.

The proposed approach does not assume agents act to either maximize their utility or to satisfy their own threshold, offering a novel perspective on the origin of segregation. The representation of agent mobility as reactions also comes with a number of advantages: (i) it utilizes the rich analytical and numerical treatment of phase separation \cite{Glotzer} in CRNs -- reminiscent of patterns observed in segregation; (ii) it allows for modeling  \textit{open} neighborhoods, i.e. those  without conservation of the number of agents of a certain type (although the number of houses is conserved); and (iii) the \emph{ergodicity} of the CRN resolves the common problem of getting stuck in a sub-optimal frozen state. Such a stochastic formulation is often adopted in a broad range of statistical-physics studies that  model collective behavior through transition rates rather than explicit optimization \cite{castellano2009statistical}.

In addition to the emergence of segregation, the open character of the model further permits a phase transition between a symmetric state, with equal representation of both agent types, and an asymmetric state in which one type dominates. This asymmetric tipping point occurs spontaneously at a critical level of in-group preference, even if the preference for like agents is symmetric between both types. By means of numerical simulations and finite-size scaling analysis, we establish the continuous, i.e. second-order, nature of this phase transition.

We note that other tipping scenarios (including both second-order transition and cross-over) have been considered previously, such  as ``type mobility'' induced by a fraction of \emph{type-switching} but \emph{non-moving} agents in 2D (regardless of their satisfaction \cite{hazan2013schelling}), or models where the entire population is able to switch their types on an 1D ring  \cite{barmpalias2015tipping}) or in a non-spatial setting \cite{kollmann2021social}. While we also use probabilistic update rules, our reactions are mediated by empty sites, since we only include transitions between blue (red) and empty sites. As discussed in \cite{boke2025varietiesschellingmodelexperience}, empty sites facilitate the ``effective'' diffusion of red and blue agents, resulting in different segregation patterns. We therefore expect significant difference between our model and the works of \cite{hazan2013schelling, barmpalias2015tipping, kollmann2021social} in terms of macroscopic behavior. Furthermore, the explicit role of vacancies resonates with sociological vacancy-chain theories of residential mobility, where empty housing units mediate population redistribution \cite{Turner}.

The rest of the article is organized as follows. In Sec.~\ref{sec:model_def}, we present the model definition in terms of microscopic processes responsible for the in-group preference and out-group tolerance. Next, we derive a mean-field theory and its second-order correction via moment closure in Sec.~\ref{sec:mean-field}. This allows us to show a cross-over in energy and a bifurcation in magnetization. The latter is responsible for the tipping point. In Sec.~\ref{sec:fss} we provide simulation results on 2D Moore neighborhoods, accompanied by finite size scaling analysis of the observables (energy, net magnetization), from which we extract the critical exponents for comparison with known universality classes. Finally, in Sec.~\ref{sec:discussion} we discuss our results in the broader context of the emerging field of social physics \cite{jusup2022social, castellano2009statistical} and statistical physics of cities \cite{barthelemy2019statistical}. All other technical details are provided in the Supplemental Material.

\section{Segregating chemical reaction network}\label{sec:model_def}

To make the connection to Schelling's original work on segregation we formulate a chemical reaction network on a two-dimensional square lattice with $N = L \times L$ total lattice sites. Each  site $i \in ([1,2, \cdots L] \otimes[1,2, \cdots L])$ can be in any of three distinct states denoted by  $A_i$, $B_i$ and $0_i$, signifying that either an agent of type $A$ or $B$, or a vacant site is present at that location on the lattice.

Next, we specify all possible reactions in our network as illustrated in  Fig. \ref{fig:overview}. 
Any change in the macroscopic state of the lattice is determined by move-in (\textit{birth}) or move-out (\textit{death}) reactions of its agents. 
These birth-death reactions can be split into (spontaneous) neighbor-independent and (interacting) neighbor-dependent processes. 
More concretely, in the first category we have,
\begin{enumerate}
    \item \textbf{Death reactions}: Agents moving out of the neighborhood with rate $\mu$:
    \begin{equation}\label{death}
        A_i  \xrightarrow{\mu} 0_i\,, \qquad 
        B_i  \xrightarrow{\mu} 0_i\,. 
    \end{equation}
    \item \textbf{Birth reactions:} Agents moving into the neighborhood with rate $\lambda$:
    \begin{equation}\label{birth}
        0_i  \xrightarrow{\lambda} A_i\,, \qquad 
        0_i  \xrightarrow{\lambda} B_i\,. 
    \end{equation}
\end{enumerate}
In addition to such a birth-death process at each lattice site $i$, we also consider interactions between $i$ and its nearest neighbors within the Moore neighborhood. The conclusions of this work do not differ if we use other neighborhoods, although the exact quantitative results will. The neighbor-dependent processes are of two types:
\begin{enumerate}
    \item \textbf{In-group reactions}: These encode the tendency of agents to attract similar agents. This happens in two ways. First, by adding move-out pressure on opposite agents with rescaled death rate $r_c \mu$. Second, by adding move-in pressure on similar agents with rescaled birth rate $r_c \lambda$. Neighboring sites $i,j \in (L,L)$ thus interact as:
    \begin{subequations}\label{copy}
    \begin{align} \label{copy1}
        0_i + A_j & \xrightarrow{r_c \lambda} A_i + A_j\,, & 
        0_i + B_j & \xrightarrow{r_c \lambda}  B_i + B_j\,, \\ \label{copy2}
        A_i + B_j & \xrightarrow{r_c \mu}  0_i + B_j\,, &
        B_i + A_j & \xrightarrow{r_c \mu}  0_i + A_j\,.
    \end{align}
    \end{subequations}
    \item \textbf{Out-group reactions}: Encode the tendency of agents to attract opposite agents. This again happens in two ways. First, by adding move-out pressure on similar agents with rescaled death rate $r_s \mu$. Second, by adding move-in pressure on opposite agents with rescaled birth rate $r_s \lambda$. Neighboring sites $i,j \in (L,L)$ thus interact as:
    \begin{subequations}\label{split}
    \begin{align} \label{split1}
        0_i + B_j & \xrightarrow{r_s \lambda} A_i + B_j\,, & 
        0_i + A_j & \xrightarrow{r_s \lambda}  B_i + A_j\,, \\ \label{split2}
        A_i + A_j & \xrightarrow{r_s \mu}  0_i + A_j\,, &
        B_i + B_j & \xrightarrow{r_s \mu}  0_i + B_j\,.
    \end{align}
    \end{subequations}
\end{enumerate}
Besides the neighborhood definition, the parameters of our model are $\mu$, $\lambda$, $r_c$ and $r_s$. Without loss of generality we can always rescale time to set one of the parameters to one, leaving three free parameters in total. 

The CRN defined by \eqref{death}-\eqref{split} is ergodic; the out-group reactions \eqref{split} are the inverse of the in-group reactions \eqref{copy}, albeit with reaction rates replaced as $\mu \leftrightarrow \lambda$ and $r_s \leftrightarrow r_c$. The ergodicity of the system makes sure that Markov chain of the CRN converges to a unique steady-state distribution. In the next section, we will use a mean-field approach to investigate how this steady state distribution depends on the model parameters. 

\section{Mean-field equations and the collective phases}\label{sec:mean-field}

To gain insight into this model, we first analyze the mean-field equations, whose solutions correspond to different collective phases of the stochastic model. The mean-field equations are readily derived from the law of mass action, i.e. from the rate equations of the CRN \eqref{death} - \eqref{split}, where we treat each node as an independent and identically distributed variable.  Alternatively, they can also be derived from the master equation of the CRN by assuming statistical independence of pair probabilities for nearest-neighbors in the lattice (see supplemental material \ref{app:mean_field}). 
The resulting equations are expressed in terms of the probabilities $\rho^X(t)$ for any site of the lattice to be in state $X \in \{A, B, 0\}$ at time $t$: 
\begin{subequations}\label{meanfieldequations}
\begin{align}
\frac{d\rho^A}{dt}
&= - T \rho^A(t) + \rho^0(t)  - 8 T \rho^A(t)
    \bigl( r_s \rho^A(t) + r_c \rho^B(t) \bigr) \nonumber \\
&\quad + 8 \rho^0(t)
    \bigl( r_c \rho^A(t) + r_s \rho^B(t) \bigr)\,,
\\
\frac{d\rho^B}{dt}
&= - T \rho^B(t) + \rho^0(t) - 8 T \rho^B(t)
    \bigl( r_s \rho^B(t) + r_c \rho^A(t) \bigr) \nonumber \\
&\quad + 8 \rho^0(t)
    \bigl( r_c \rho^B(t) + r_s \rho^A(t) \bigr)\,,
\end{align}
and, owing to the vacancy constraint on each site:
\begin{equation}
\rho^0(t)  = 1 - \rho^A(t) - \rho^B(t)\,.
\end{equation}
\end{subequations}
Here we have rescaled time $\tilde{t} = \lambda t$ and defined $T = \mu/\lambda$ for notational convenience. 
The probabilities $\rho^X(t)$ are equivalent to mean density of type $X$ on the lattice.

The mean-field equations \eqref{meanfieldequations} have four steady-state solutions. One is always unphysical, giving negative probabilities. 
Whenever,
\begin{equation}\label{eq:symmcondition}
    r_c < r_c^* \equiv \frac{1}{4} + 3 r_s + \frac{T}{8}
\end{equation}
there is a single physical solution to the mean-field equations (the others give negative or imaginary probability densities). In this state the probabilities $\rho^A$ and $\rho^B$ are exactly equal, and there is no net overdensity on the lattice. For this reason we call the phase associated to this state the \textit{symmetric phase}, with the associated densities:
\begin{equation}\label{eq:symmphase}
    \rho^0 = \frac{T}{2+T} \,, \quad \rho^A = \frac{1}{2+T} = \rho^B
\end{equation}
When condition~\eqref{eq:symmcondition} is not met, the symmetric solution becomes linearly unstable, while two other stable branches of solutions become physical (i.e. real-valued and bounded $\in [0,1]$). The system will settle into one of these branches, depending on initial conditions. On both branches, the $\mathbb{Z}_2$ symmetry exchanging $A \leftrightarrow B$ is explicitly broken, as one type of agent becomes more abundant. Since here $\rho^A \neq \rho^B$ we call this the \textit{asymmetric phase}. The densities in this phase are given by:
\begin{subequations}\label{assymetric_steadystate}
    \begin{align}
        \rho^A & = \frac{8\left(r_c - r_s\right) - T \pm \sqrt{D}}{16\left(r_c + r_s\left(T-1\right)\right)} \\
        \rho^B & = \frac{8\left(r_c - r_s\right) - T \mp \sqrt{D}}{16\left(r_c + r_s\left(T-1\right)\right)}\\ 
        \rho^0 & = \frac{T\left(r_s + 1/8\right)}{r_c + r_s\left(T-1\right)} 
    \end{align}
\end{subequations}
where,
\begin{equation}
    D = (-2 + 8 r_c - 24 r_s - T) (2 + 8 r_c + 8 r_s - T)
\end{equation}
Note that the two branches are the same solution with the densities for $\rho^A$ and $\rho^B$ switched. In analogy with spin models, the symmetric and asymmetric phases of our model correspond to disordered and ordered phases of a magnetic system. Therefore, we can use $M(t) = \rho^A(t) - \rho^B(t)$ as an order parameter to quantify the overdensities of one type of agent. In the asymmetric phase, there is a net overdensity $M(t\rightarrow \infty)\neq 0$ at stationary:
\begin{equation}
    |M| = |\rho^A - \rho^B| = \frac{\sqrt{D}}{8\left(r_c + r_s\left(T-1\right)\right)} \,,
    \label{overdensity}
\end{equation}
which implies that one of the agent types will start to dominate in the neighborhood. Which one depends on the initial conditions in the deterministic mean-field model, or on fluctuations in the stochastic model.

\begin{figure}
\includegraphics[width=\linewidth]{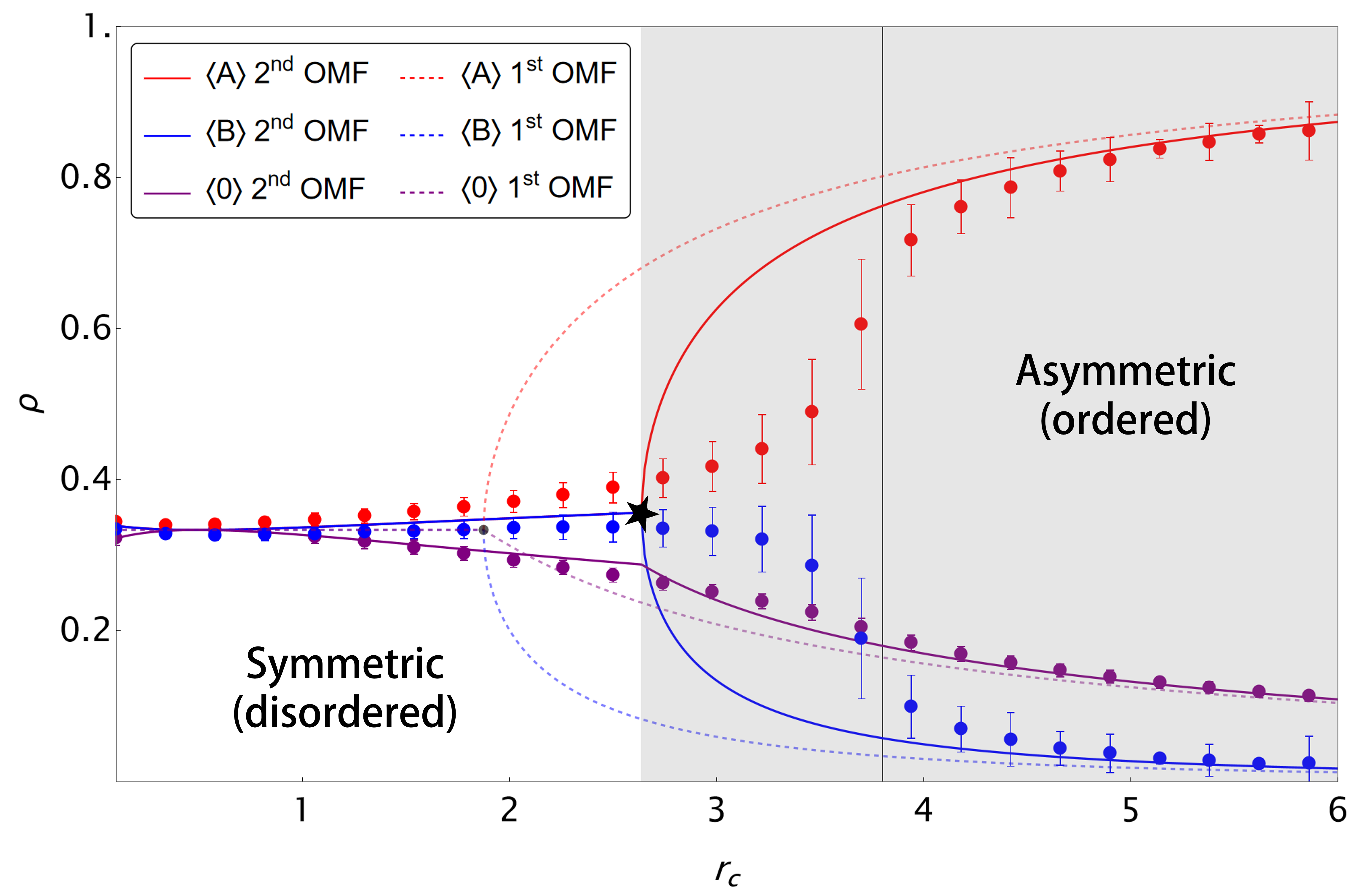}
\caption{\label{fig:mfsolutions} The densities in the stable steady states for $T = \mu /\lambda = 1$ and $r_s = 0.5$, as a function of $r_c$, obtained from the second-order (full lines) and first-order (dashed lines) mean-field approximations. We show the density of vacant sites (purple), of agents of type $A$ (red) and of agents of type $B$ (blue). The scatter data corresponds to the averages of $10^4$ independent instances of the stochastic model, with grid-size $N = 50 \times 50$. The mean-field analysis shows a  pitchfork bifurcation at a critical value  $r_c$ (marked by a black star), where the solution changes from  symmetric to asymmetric one. The mean field solutions agree with the stochastic model away from $r_c^*$ (marked by a vertical line $r_c \approx 3.8$). Note that there is also an asymmetric solution for which the densities of $A$ and $B$ are flipped with respect to the densities shown here.}
\end{figure}

The densities $\rho^A, \rho^B$ and $\rho^0$ as predicted by the (first-order) mean-field approximation are plotted as dashed lines in Fig.~\ref{fig:mfsolutions}. Here we also compare the mean-field results with expectation values obtained from numerical simulations of the stochastic model for a grid of size $N = 50 \times 50$.  We confirm two distinct phases, the disordered (symmetric) and the ordered (asymmetric) ones, which are separated by a pitchfork bifurcation at the critical point.  As expected, away from the critical point, mean-field predictions are accurate, but long-range correlations spoil their validity when $r_c \sim r_c^*$. In order to improve upon the first-order mean-field theory, we have performed a second-order moment closure \cite{Wuyts2022,kuehn2024preserving} in the supplemental material Sec. \ref{sec:2ndorderMF}. As shown by the solid lines in Fig.~\ref{fig:mfsolutions}, the second-order moment closure significantly improves the estimation of $r_c^*$, although it still deviates from the numerical value. The second-order moment closure qualitatively agrees with the stochastic model.

Besides the tipping point at $r_c=r_c^*$, we identify an additional regime when interpreting the model in socioeconomic terms. Specifically, whenever $r_s > r_c$, lattice sites in state $A$ are more likely to be neighbored by lattice sites in state $B$ and vice versa. This corresponds to a type of anti-ferromagnetic configuration, which can be interpreted as agents' preference for having diverse neighborhoods. In what follows we refer to this phase as \textit{anti-segregating}. Moreover, we observe that a global measure of the net abundance of like neighbors, defined in a similar fashion to the (Ising) energy, passes through a zero at $r_c = r_s$, when the system transits from segregating to anti-segregating patterns. However, as seen in the next section, such a transition is not critical, but rather a simple cross-over from negative to positive energy.

\begin{figure}
\includegraphics[height=0.9\linewidth]{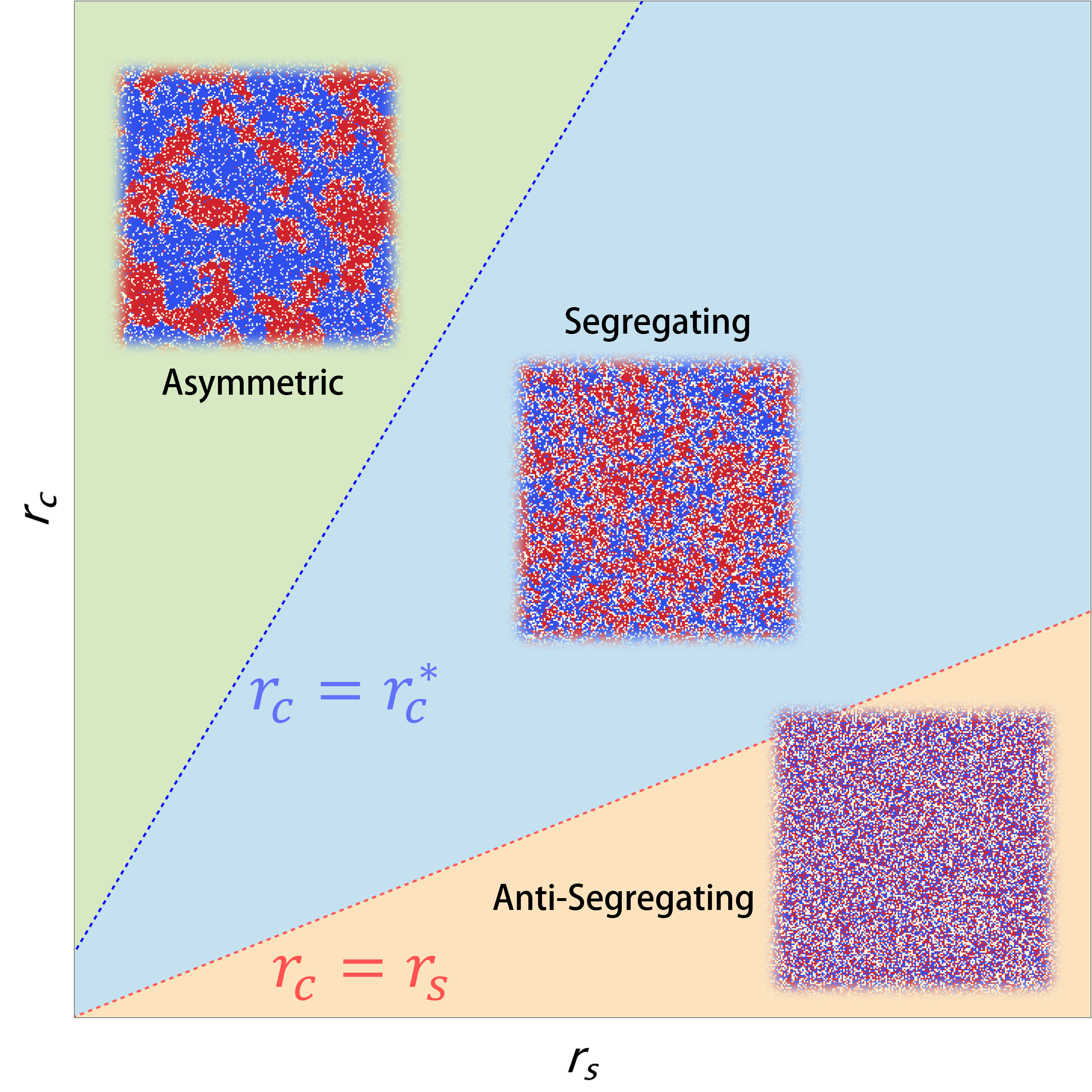}
\caption{\label{fig:phasediagram} Schematic phase diagram of our model in the $(r_s, r_c)$-plane. The symmetric and asymmetric phases are separated by a \textit{tipping} transition at $r_c^*$, while a cross-over transition from anti-segregating to segregating neighborhoods takes place when $r_c = r_s$. The inlay shows typical configurations for each phase, obtained from simulations of the model using Gillespie's algorithm.}
\end{figure}

In the remainder of the paper, we will analyze the dynamics and critical behavior of the asymmetric tipping point using exact stochastic simulations (Gillespie's algorithm \cite{gillespie1977exact}). Details on the simulation algorithm are given in the supplemental material (SM) Sec.~\ref{sec:appendix_implementation} and a comparison between the mean-field solution and a stochastic simulation run is detailed in the SM Sec. \ref{sec:comparison}. We show the mean-field phase diagram of the model in Fig. \ref{fig:phasediagram}, along with typical configurations from each of the three phases.
The transitions displayed in the figure are:
\begin{enumerate}
    \item A cross-over from an \textit{anti-segregating} neighborhood to a \textit{segregating} one. Here we define \textit{anti-segregating} as a neighborhood in which agents of opposite types are more likely to be neighbors than not. A \textit{segregating} neighborhood represents the opposite situation. This transition is characterized by a cross-over from negative to positive energy, where the energy is defined as the relative abundance of like neighbors.
    \item The asymmetric \emph{tipping} transition from a neighborhood with an (approximately) equal number of agents of each type, to one in which one of the two types gains the upper hand. This transition is akin to the transition from a paramagnetic to a ferromagnetic phase in the Ising model, characterized by a shift from zero magnetization to nonzero net magnetization and a peak in the variance of the energy. From a sociological point of view this transition is ``maximally'' segregating as the whole neighborhood starts to be dominated by one type of agent. 
\end{enumerate}
Of these two transitions, only the second one is critical, in the sense that the correlation length $\xi$ diverges at the critical point in the thermodynamic limit.

\begin{figure}
    \centering
    \includegraphics[height=0.9\linewidth]{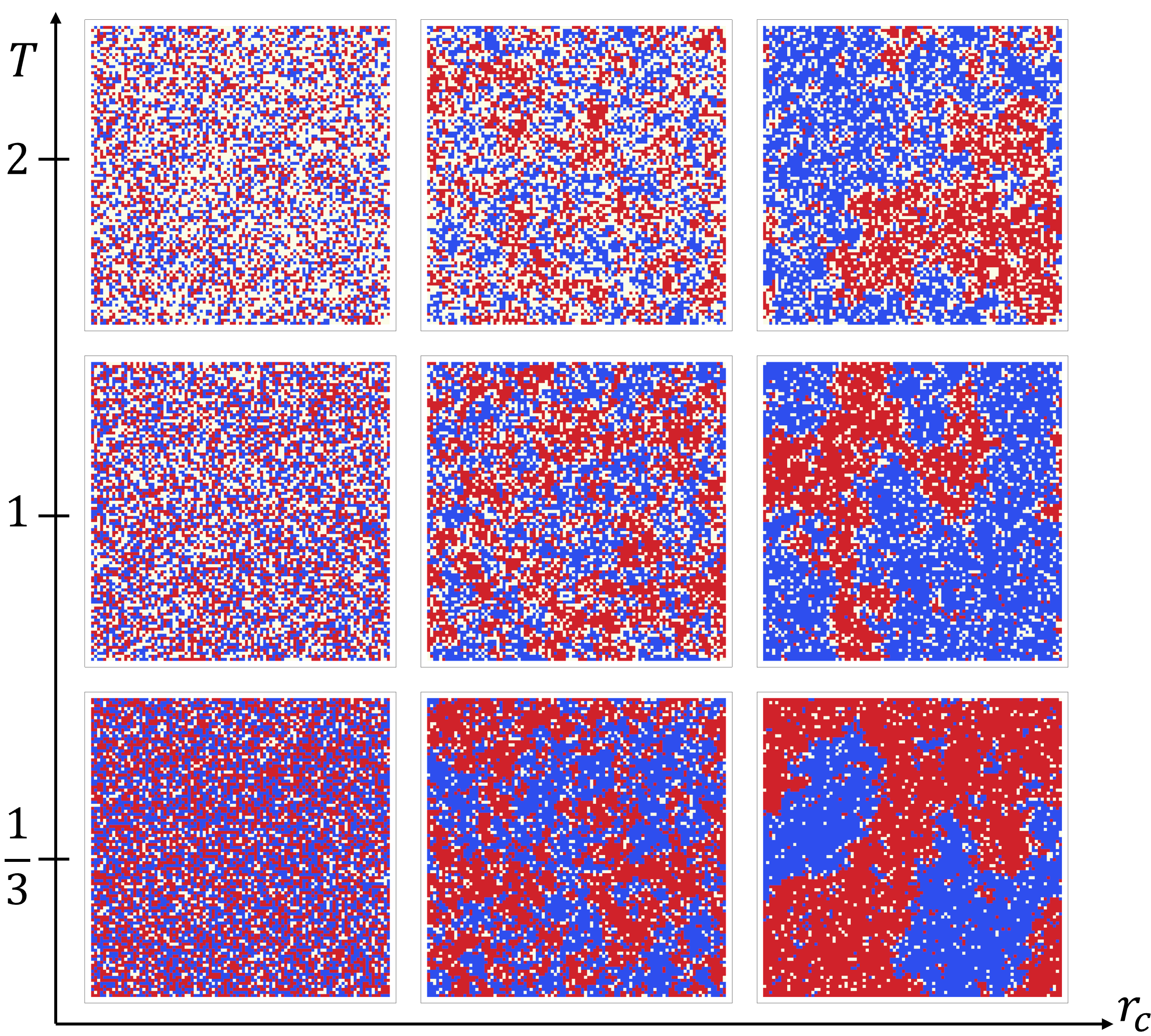}
    \caption{Snapshots of steady state solutions of the stochastic model obtained by varying $T = \mu/\lambda$ and $r_c$ independently while fixing $r_s$ (here $r_s = 0.5$). Note that this plot is merely meant for schematic purposes, the axes are not to scale.}
    \label{fig:Tvsrcgrid}
\end{figure}

We conclude this section with a short discussion on varying the effective move-out (death) rate $T = \mu/\lambda$. With our choice of parameters, we have intentionally made sure that $T$ controls to overall density of agents, and does not affect the qualitative behavior of the model. The dependence of the stochastic model on $T$ is captured quite accurately by the mean-field solution, confirming our understanding that $T$ is an overall tuning parameter of the model. As shown in Fig. \ref{fig:Tvsrcgrid}, increasing $T$ lowers the density of agents of both types, while preserving the three phases described above. 

\section{Phase transitions, Finite Size Scaling and Criticality}\label{sec:fss}
To further characterize the model and investigate its universality class we will now find the critical exponents of the asymmetric tipping transition. To do so, we use the well-established finite size scaling method. In what follows we briefly introduce this method before discussing the results for our chemical reaction network.

\subsection{Finite Size Scaling}
At a continuous phase transition the correlation length $\xi$ of an infinite system diverges. Therefore, at fixed $\mu$, $\lambda$ and $r_s$, we expect our system near the critical point $r_c^*$ to have
\begin{equation}
    \xi \sim |r_c - r_c^*|^{-\nu}
    \label{eq:corrlengthcrit}
\end{equation}
where $\nu$ is the critical exponent for the correlation length. An observable $O$ scaling with a critical exponent $\zeta$ near the transition thus depends on the correlation length as,
\begin{equation}
    O \sim |r_c - r_c^*|^{-\zeta} \sim \xi^{\zeta/\nu}\,.
    \label{eq:criticalobservable}
\end{equation}
In our simulations with lattices of finite size $N = L\times L$, typically, $L \ll \xi$, so the system size effectively bounds the maximum correlation length. In this regime we conclude,
\begin{equation}
    O \sim L^{\zeta/\nu}
    \label{eq:critobslargeL}
\end{equation}
This logic leads to the finite size scaling Ansatz, which holds near the phase transition:
\begin{equation}
    O \sim L^{\zeta/\nu} f\left(L^{1/\nu}(r_c - r_c^*)\right)\,.
    \label{eq:FSSansatz}
\end{equation}
Here $f(x)$ is a universal scaling function determining how finite size effects influence the value of observables. By our previous discussion, we know that $f(x) \rightarrow \mathrm{const.}$ as $x \rightarrow 0$, and that $f(x) \rightarrow x^{-\zeta}$ as $|x|\rightarrow\infty$. The finite-size Ansatz in Eq.~\eqref{eq:FSSansatz} allows us to determine the critical exponents of observables. In particular, the correct values of the exponents uniquely collapse the data of $L^{-\zeta/\nu} O$ plotted against $L^{1/\nu}(r_c - r_c^*)$ onto each other for different system sizes. To do this effectively we need a way to determine the value of $r_c^*$. We can do this independently by using the Binder cumulant. 

The results shown in this section are obtained by performing numerical simulations of our stochastic model using a Gillespie-type algorithm. We start by initializing the lattice randomly and evolving for 180 Monte Carlo (MC) sweeps to find an approximate steady state for a given set of parameters. We compute expectation values of observables by taking measurements every subsequent 18 MC steps (to avoid a strong auto-correlation) and averaging over a total of $n_{\rm sim}$ such measurements.

\subsection{Binder cumulant}
To make the analogy with spin models explicit (as originally suggested for Schelling's model in \cite{gauvin2009phase}) we link the overdensity $M$ defined in Eq.~\eqref{overdensity} with the microscopic (stochastic) variables
\begin{equation}
    M = \frac{1}{N}\sum_i \sigma_i
    \label{eq:magnetization}
\end{equation}
where $M$ represents the net ``magnetization'' on the lattice, and $\sigma_i = 1,0,-1$, if site $i$ is occupied by agent $A$, empty or occupied by agent $B$ respectively.
Another interesting observable is the average energy on the lattice,
\begin{equation}
    E = \frac{1}{4 N}\sum_{\langle i,j\rangle} \sigma_i\sigma_j
    \label{eq:energy}
\end{equation}
which is a measure of how segregated a neighborhood is locally, as it measures the net discrepancy between neighbors of like and opposing type. While its definition suggests that $E$ is related to the energy in pairwise interacting spin models, it is not used here to specify the dynamical evolution of the system, but only as a measure of segregation. Furthermore, when measuring this energy, we use the Von Neumann (left, right, up, down) neighborhood, conventional in Ising-like spin systems.

\begin{figure}
    \centering
    \includegraphics[width=0.8\linewidth]{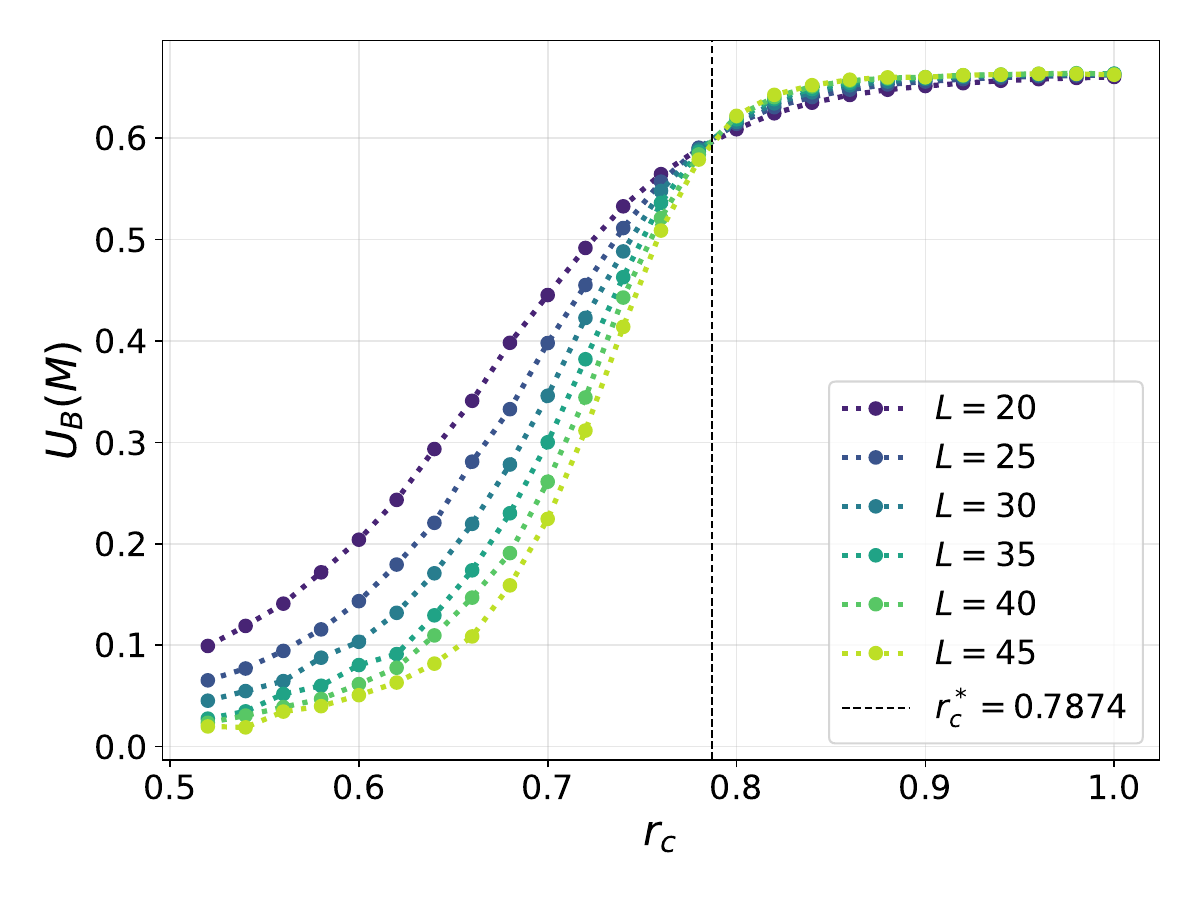}
    \caption{The Binder cumulant for the magnetization characterizes the continuous phase transition to the asymmetric (ferromagnetic) phase where agents of one type become the dominant group in the region. The critical value, shown here for $r_s =0$, is determined at the crossing point for the Binder cumulants as $r_c^* = 0.7874(2)$.}
    \label{fig:Bindercumulant}
\end{figure}

To determine the exact value of $r_c^*$ we use the Binder cumulant of the magnetization,
\begin{equation}
    U_B(M) = 1 - \frac{\langle M^4\rangle}{3\langle M^2\rangle^2}
    \label{eq:binder}
\end{equation}
which goes to a scale independent universal value at the critical value $r_c^*$. In Fig. \ref{fig:Bindercumulant} we plot $U_B(M)$ for different system sizes $N = L \times L$ and determine $r_c^*$ by identifying the crossing points of all the curves. The value for the Binder cumulant at the crossing is found to be $U_B(M)|_{r_c^*} = 0.60$, which is close to that of the Ising model (there $U_B(M)|_{r_c^*} = 0.61$ with periodic boundary conditions \cite{selke2006critical}).

\begin{figure}
    \centering
    \includegraphics[width=0.8\linewidth]{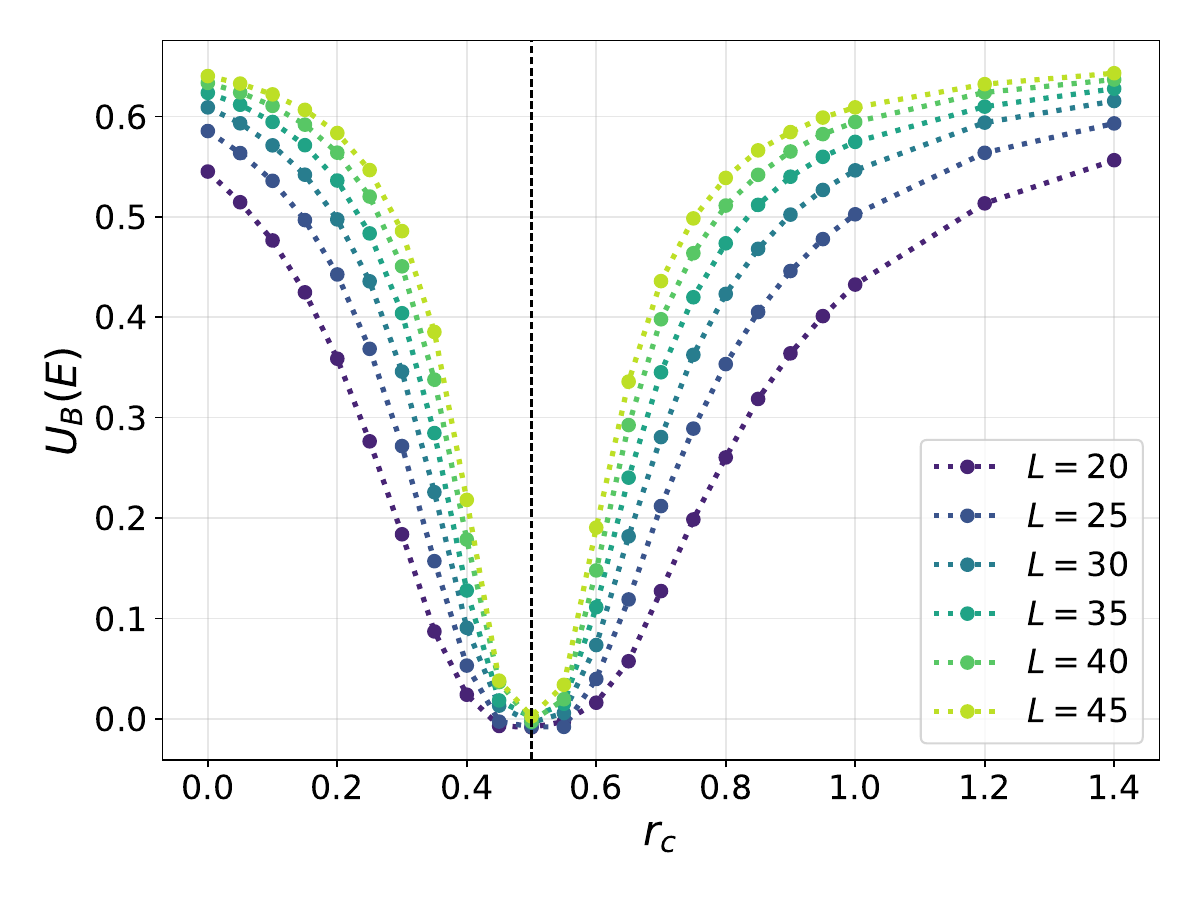}
    \caption{The Binder cumulant for the energy for $r_s = 0.5$ and $T=\mu/\lambda = 1$ dips to zero at $r_c = r_s$. Dotted line is to indicate the value of $r_s$.}
    \label{fig:EnergyBindercumulant}
\end{figure}

Earlier, we characterized the transition from anti-segregating to segregating as a smooth cross-over occurring at $r_c = r_s$, corresponding to the absence of a net preference in neighbor type. At this point the average energy defined in Eq.~\eqref{eq:energy} changes sign, as shown in Fig.~\ref{fig:secondorderMF2}. Since this cross-over is not associated with a diverging correlation length, the energy fluctuations remain approximately Gaussian in shape around this point. Phenomenologically,
\begin{equation}
    P(E) \sim \frac{1}{\sqrt{2 \pi \sigma^2}} 
    \exp\!\left[-\frac{(E-\langle E\rangle)^2}{2\sigma^2}\right],
    \label{eq:pofe}
\end{equation}
which implies a vanishing Binder cumulant $U_B(E)=0$ for a perfectly Gaussian distribution. This behavior is observed numerically in Fig.~\ref{fig:EnergyBindercumulant}, where the Binder cumulant exhibits a pronounced dip near $r_c=r_s$, independently of system size. Away from this point the distribution becomes sharply peaked such that $U_B(E)$ approaches $2/3$ for increasing system sizes.

We emphasize that the near-Gaussian form of $P(E)$ at the cross-over does not preclude critical scaling of energy fluctuations at the true critical point $r_c=r_c^*$. In particular, while the shape of the distribution remains close to Gaussian, its variance $\mathrm{Var}(E)$ can still exhibit system-size scaling, giving rise to critical behavior in the heat-capacity. In summary, the Binder cumulant of the energy probes the Gaussian character of fluctuations at the cross-over, whereas the Binder cumulant of the magnetization displays the standard crossing behavior characteristic of a continuous phase transition.

\subsection{Critical exponents}
\begin{figure*}
    \centering
    \includegraphics[width=0.5\columnwidth]{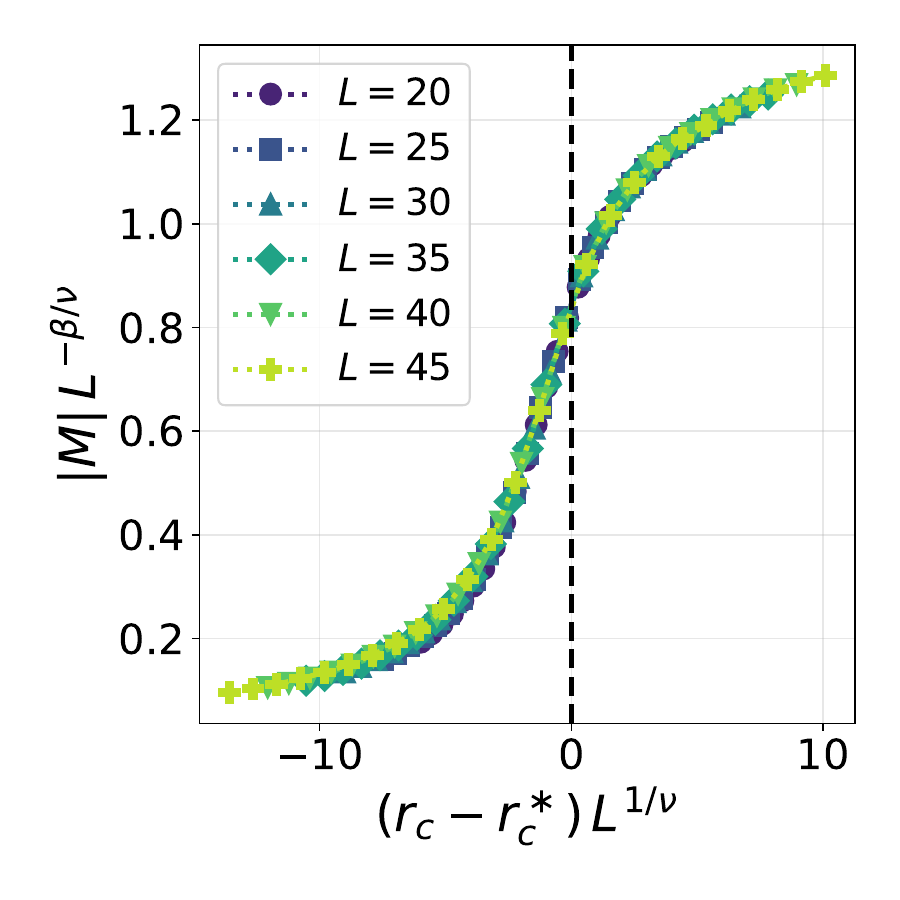}
    \includegraphics[width=0.5\columnwidth]{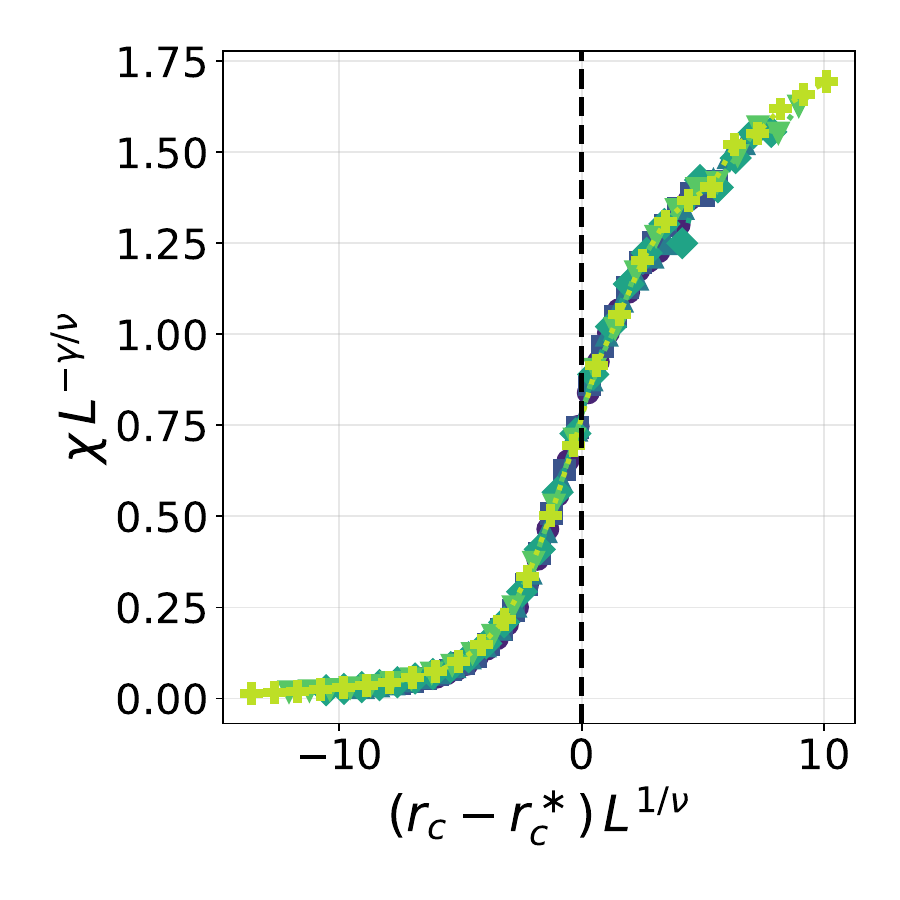}
    \includegraphics[width=0.5\columnwidth]{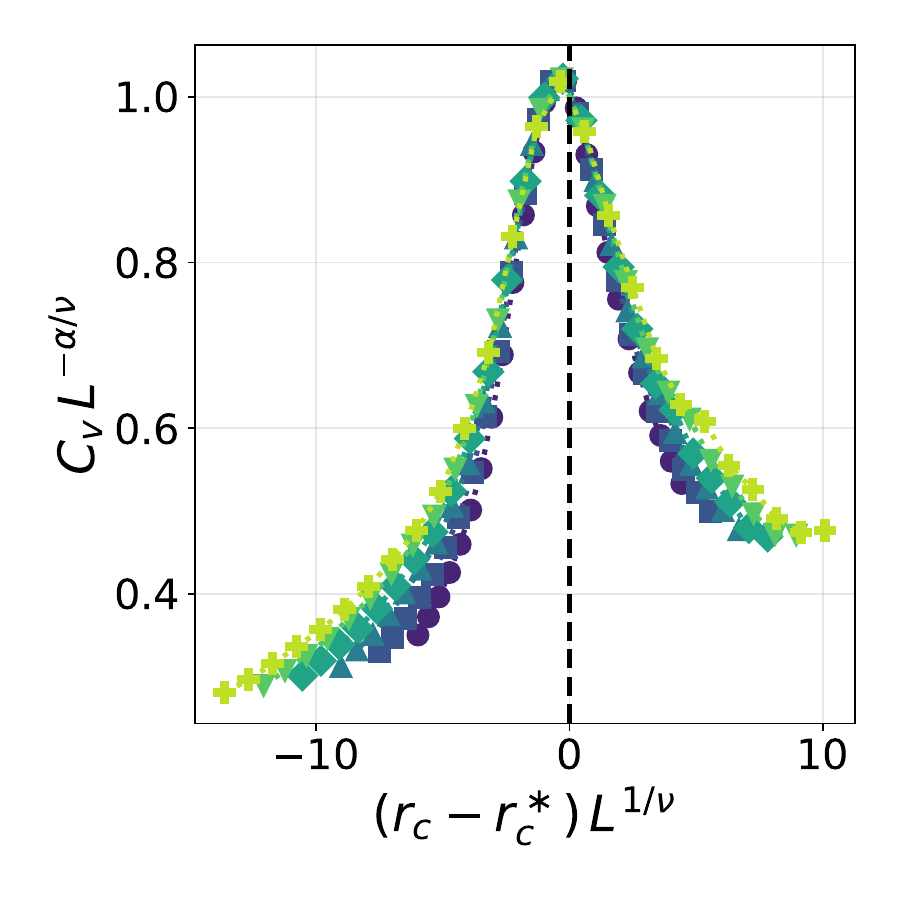}
    \includegraphics[width=0.5\columnwidth]{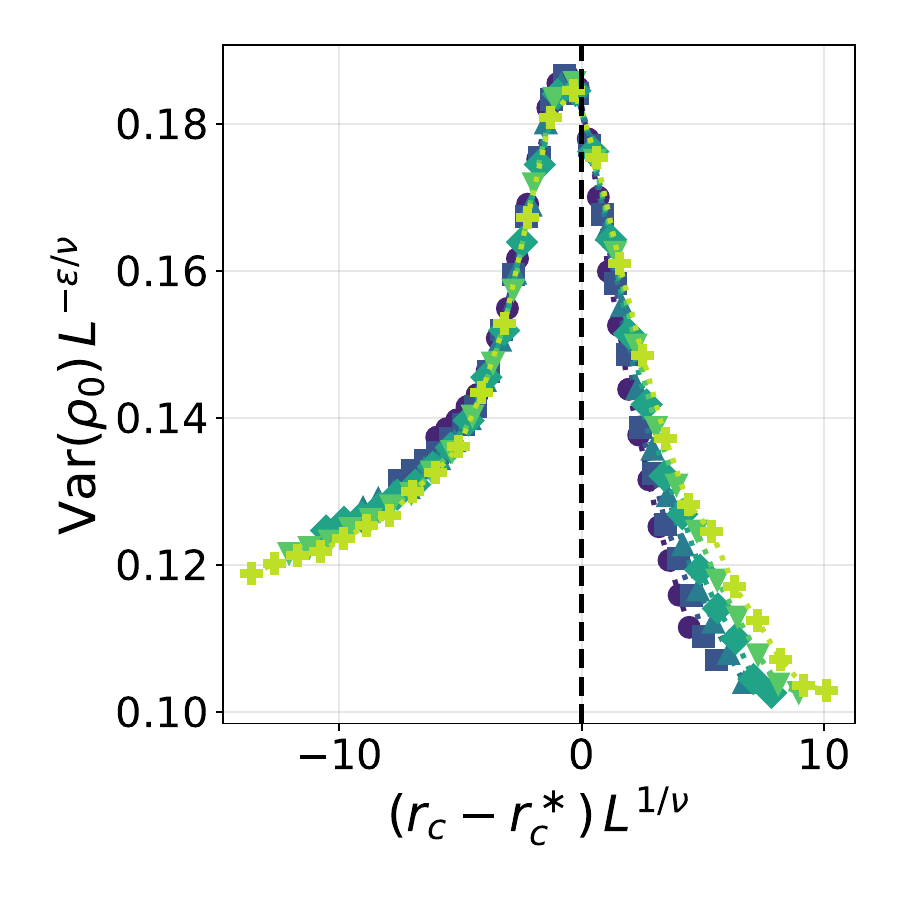}
    \caption{Rescaling observables by their exponents at $r_c^*$ shows a collapse of the curves onto a single line, shown here at $r_s=0$ for the observables $|M|, \chi, C_V$ and ${\rm Var}(\rho_0)$, respectively. The value of $\nu$ was estimated by determining the best overlap for $|M|L^{-\beta/\nu}$, fixing it for the other three graphs.}
    \label{fig:scaling_collapse}
\end{figure*}

Now that we have determined $r_c^*$ belonging to the infinite system size limit, we can use the finite scaling Ansatz to determine the value of the critical exponents for various observables. We focus here on the scaling of the absolute magnetization per site $|M| \sim L^{\beta/\nu}$, the susceptibility, computed as:
\begin{equation}
    \chi = L^2 (\vev{M^2} - \vev{M}^2 )\,,
\end{equation}
and its exponent $\gamma$. Additionally, we characterize two more exponents at the critical point ($\alpha$ and $\epsilon$) by determining the critical scaling of the variance in energy (or, the heat capacity):
\begin{equation}
    C_V =L^2( \vev{E^2} - \vev{E}^2)\,,
\end{equation}
and the variance in the density of vacant sites ${\rm Var}(\rho_0)$, respectively. After determining the scaling behavior at the critical point, we can determine $\nu$ using \eqref{eq:FSSansatz} for the magnetization $|M|$. In practice, this boils down to rescaling the $r_c$ axis as $(r_c-r_c^*)L^{1/\nu}$ and optimizing for the $\nu$ which minimizes the variance in $|M| L^{-\beta/\nu}$ for various system sizes, effectively `collapsing' all magnetization curves on top of each other. This is shown for the observables of interest in Fig.~\ref{fig:scaling_collapse}. We perform this analysis for three values of $r_s$ (shown in Table~\ref{tab:criticalexponents}) while varying $r_c$ and $N$. The number of independent simulation runs $n_{\rm sim}$ for each $(r_c, r_s, N)$-choice is shown in the bottom row of the table. The error margins in the reported critical exponents are obtained by the bootstrapping method, where 100 datasets where created by sampling with replacement from the obtained data and the finite size scaling analysis was repeated for each of these datasets. Note that this analysis omits other types of systematic errors, which may arise from the finite system size window chosen here ($L=$ 20-45) or possible insufficient equilibration near the critical point resulting in the underestimation of the autocorrelation.

\begin{table}
    \centering
    \begin{tabular}{ccccc}
        Observable & $\zeta$ & $r_s = 0$ & $r_s = 0.1$ & $r_s = 0.5$\\
        \hline
        $r_c^*$ & & $0.7874(2)$ & $1.3870(4)$  & $3.8039(16)$\\
        $|M|$ & $\beta$ &  $-0.1407(8)$ & $-0.1420(8)$ & $-0.1295(14)$ \\
        $\chi$ & $\gamma$ & $1.8766(15)$ & $1.9013(16)$ & $1.8693(55)$\\
        $C_V$ & $\alpha$ & $0.2937(35)$ & $0.2967(35)$ & $0.2845(65)$ \\
        ${\rm Var}(\rho_0)$ & $\epsilon$ & $-1.975(4)$ & $-2.003(4)$ & $-1.944(10)$ \\
        $\xi$ & $\nu$ & $1.0816(9)$ & $1.0928(5)$ & $1.0660(31)$ \\
        $n_{\rm sim}$ & & $256\,000$ & $200\,064$ & $128\,000$
    \end{tabular}
    \caption{Critical exponents for several observables, assuming $O \sim L^{\zeta/\nu}$ for the exponent $\zeta$. Uncertainties are reported as the standard deviation over 100 exponents, obtained by resampling the simulated outputs with replacements (i.e. bootstrapping).}
    \label{tab:criticalexponents}
\end{table}

The ratios of critical exponents such as $\beta/\nu$ and $\gamma/\nu$ are close to the Ising values, suggesting that the transition is governed by a $\mathbb{Z}_2$ symmetry-breaking mechanism akin to Ising-like criticality. This feature is consistent with a broad class of non-equilibrium systems in which Ising universality emerges despite the absence of detailed balance \cite{de1993nonequilibrium,odor2004universality}. From this perspective, the non-equilibrium nature of the CRN dynamics does not, by itself, preclude Ising-like criticality. However, the measured critical exponents in Table~\ref{tab:criticalexponents} differ from those of the standard 2D Ising and Potts universality classes. In particular, the heat capacity shows a power-law divergence, rather than the logarithmic behavior characteristic of the 2D Ising model. In addition, the independently measured exponents do not satisfy standard hyperscaling relations when interpreted within an equilibrium framework. These deviations indicate that the critical behavior of the present model cannot be straightforwardly identified with that of a standard equilibrium universality class such as the Ising or Potts models.

To assess whether these conclusions are affected by the chosen range of  $L$, in Appendix~\ref{app:finite-size-robustness} we performed additional  robustness analysis for moderate system sizes $L=50,55,60$. In brief, we repeated the full finite-size-scaling procedure for a family of fitting windows, recomputing the Binder estimate of $r_c^*$ and the critical exponents independently for each window. This included both an upper-cutoff scan, in which the maximum system size was varied as $L_{\max}=45,50,55,60$, and a lower-cutoff scan, in which the smallest system sizes were successively removed. Across these windows, the estimates of $\beta/\nu$ and $\gamma/\nu$ remain stable within the statistical uncertainty of the bootstrap analysis. The heat-capacity exponent $\alpha/\nu$ shows the largest window dependence, as expected for an energy-like observable with stronger finite-size corrections, but it remains incompatible with a purely logarithmic Ising-like divergence over the accessible range of system sizes. The collapse estimate of $\nu$ also varies only moderately under changes of the fitting window. We therefore find no evidence that the deviations from 2D Ising-universality class reported in Table~\ref{tab:criticalexponents} can be removed by extending the analysis from $L\leq45$ to $L\leq60$.

One possible interpretation is that these deviations arise from strong corrections to scaling, which may still be significant even over the extended system-size range. Another possibility is that the critical behavior is influenced by an additional coupled field, namely the local vacancy density, which plays an explicit dynamical role in the CRN formulation. Although the vacancy density is not globally conserved in the present model, it is constrained at the single-site level and is dynamically coupled to the order parameter. Such coupling may lead to effective renormalization of scaling behavior, particularly if density fluctuations evolve on time scales comparable to those of the order parameter near criticality \cite{hohenberg1977theory,folk2003critical}.

In equilibrium statistical mechanics, analogous situations are known to give rise to renormalized critical behavior when thermodynamic constraints or hidden variables couple to the energy-like scaling field \cite{fisher1968renormalization}. Whether a similar mechanism is operative in the present non-equilibrium setting remains an open question. Establishing such a connection would require demonstrating that density fluctuations act as an effective constraint or slow mode at criticality, for instance through time-scale separation, ensemble comparisons, or controlled modifications of the dynamics that fix or suppress density fluctuations.

\section{Discussion}\label{sec:discussion}

In this work, we have proposed and analyzed a non-equilibrium chemical reaction network (CRN) model of segregation with macroscopic pattern formation resulting from interactions between agents. Unlike traditional approaches, the present model does not rely on utility maximization, explicit thresholds, or deterministic relocation rules. Instead, segregation emerges from purely stochastic, nearest-neighbor reaction processes in the Moore neighborhood of a square lattice. 
We have shown that, beyond a critical level of in-group preference, the system undergoes a continuous tipping transition. This transition spontaneously breaks the $\mathbb{Z}_2$ symmetry between the two agent types and shows diverging correlation lengths and susceptibilities. Finite-size scaling analysis yields a consistent set of critical exponents describing this transition. 

The observed phenomenology is qualitatively reminiscent of Ising-like critical behavior: the order parameter is non-conserved, interactions are short-ranged, and exponent ratios such as $\beta/\nu$ and $\gamma/\nu$ are close to their two-dimensional Ising counterparts. Quantitatively, however, the measured thermal exponents $\nu$ and $\alpha$ deviate from their equilibrium Ising values, and the heat capacity exhibits a clear power-law divergence rather than the logarithmic behavior expected in the 2D Ising model. Taken together, these observations indicate that while the present model shares important features with Ising universality at the level of symmetry and exponent ratios, the full set of measured exponents cannot be straightforwardly assigned to a known equilibrium universality class. We have tested the robustness of this conclusion by extending the finite-size-scaling analysis to $L=50,55,60$ and by repeating the exponent extraction over multiple sliding fitting windows; see Appendix~\ref{app:finite-size-robustness}. This analysis shows that the main exponent ratios remain stable under changes of the fitting window, while the strongest residual finite-size dependence appears in the heat-capacity exponent. Thus, although corrections to scaling cannot be ruled out, the extended analysis does not indicate a drift toward the standard two-dimensional Ising values within the accessible system-size range. Possible explanations therefore include strong corrections to scaling persisting beyond the simulated sizes, or the presence of an additional coupled field---such as the vacancy density---that modifies the effective scaling behavior. Disentangling these scenarios requires further investigation, in particular, a systematic analysis of finite-size effects and the transient density fluctuations near criticality.

Several other directions for future research naturally arise from the present work. 
While our analysis demonstrates a systematic improvement of the analytical description through the inclusion of higher-order moment closures (see supplemental material~\ref{app:mean_field}), a general closed-form characterization of the steady-state distribution remains an open problem. One possible avenue is to explore approaches based on nonlinear graph Laplacians, whose minimizers have been shown to capture segregation-like structures in related settings \cite{POLLICOTT}.

Another important direction concerns the role of network structure in shaping the nature and location of the tipping transition \cite{banos2012network,avetisov2018phase}. Extending the present analysis to heterogeneous networks using degree-based mean-field theory and higher-order moment closures could provide analytical insight into how connectivity, degree fluctuations, and spatial structure influence segregation dynamics and critical behavior. Beyond stationary properties, it would also be of interest to study dynamical aspects of the transition, such as the statistics of tipping times or the time required for a minority group to become dominant. These questions can be naturally formulated as first-passage time problems and, in suitable limits, may be addressed through mappings to Schr\"odinger-type equations, as done for variants of the voter model in \cite{vazquez2004ultimate}.

It is worth emphasizing that the CRN framework adopted here admits a continuous time Markov chain (CTMC) description in terms of transition probabilities between global configurations. In contrast to most CTMC-based models of social dynamics \cite{rogers2011unified}, these transition rates do not rely on predetermined thresholds or explicit utility functions \cite{grauwin2009competition,GRAUWIN_Game,barbaro2013territorial}. This feature opens the possibility of incorporating state-dependent utilities or decision rules directly into the reaction rates, thereby providing a natural bridge between the present stochastic formulation and utility-based or hydrodynamic approaches to social dynamics \cite{seara2023sociohydrodynamics,zakine2024socioeconomic,Garnier_Brun}.

More broadly, the CRN formulation offers a promising route toward data-driven modeling of segregation dynamics. Because the model is expressed in terms of reaction rates rather than latent individual utilities, it is possibly more directly comparable to empirical data. In an urban context, these rates can be interpreted as coarse-grained propensities for residential mobility, local demographic turnover, or changes in neighborhood composition. National statistics bureaus, population registers, census agencies, and administrative records often provide data on residential mobility across urban regions and, in some cases, at the national level, together with demographic information on households and neighborhoods. Such data could be used to estimate effective transition rates between local population states, for example by applying Bayesian parameter-inference methods for continuous-time Markov models. This would make it possible to test whether observed local demographic changes are adequately described by uncorrelated birth--death processes, or whether models including nearest-neighbor interactions of the type proposed here provide a better statistical fit to empirical residential dynamics.

Another natural direction concerns the spatial substrate on which the dynamics takes place. In the present work the model is defined on a regular two-dimensional lattice, which is chosen for its convenience in studying local interactions, symmetry breaking, and systematic finite-size scaling. Real urban systems, however, are not regular Euclidean grids: their geometry is shaped by street networks, land-use constraints, transport infrastructure, and historical development, and often displays heterogeneous or fractal-like spatial structure \cite{batty1994fractal,batty2013new,louf2014scaling,murcio2015multifractal,merbis2026hyperscaling}. Extending the CRN model to irregular urban networks, weighted graphs, or fractal-like geometries would therefore be an important step toward more realistic applications. In such extensions, lattice sites could represent neighborhoods, census tracts, or spatial cells, while edges would encode adjacency, mobility flows, commuting connections, or other forms of spatial interaction. This would allow one to investigate how  the heterogeneous connectivity and multi-scale geometry characteristic of real cities modify the critical behavior and segregation patterns identified here.

Finally, it would be interesting to explore the interplay between the residential dynamics considered here and additional social processes such as type-switching or opinion dynamics, including voter-like models \cite{Redner2019, pham2025polarisation} or state-swapping dynamics \cite{domic2011dynamics}. Previous studies, including Schelling–voter hybrids \cite{caridi2017characterizing,feliciani2017and,ferri2023three}, have shown that such couplings can lead to rich and nontrivial pattern formation, suggesting a wide landscape of collective behaviors yet to be explored within the CRN framework.

\begin{acknowledgments}
We thank PK Mohanty for his  valuable comments.
WM is supported by the NWA ORC programme \emph{Emergence at all scales}. TMP was supported by the Dutch Institute for Emergent Phenomena (DIEP) cluster at the
University of Amsterdam under the Research Priority Area \emph{Emergent Phenomena in Society: Polarisation, Segregation and Inequality}.
\end{acknowledgments}


%


\clearpage
\onecolumngrid

\setcounter{section}{0}
\setcounter{equation}{0}
\setcounter{figure}{0}
\setcounter{table}{0}

\renewcommand{\thesection}{S\arabic{section}}
\renewcommand{\theequation}{S\arabic{equation}}
\renewcommand{\thefigure}{S\arabic{figure}}
\renewcommand{\thetable}{S\arabic{table}}

\makeatletter
\renewcommand{\theHequation}{S\arabic{equation}}
\renewcommand{\theHfigure}{S\arabic{figure}}
\renewcommand{\theHtable}{S\arabic{table}}
\makeatother

\makeatletter
\begin{center}
  \Large
  Supplemental Material for\\[0.5em]
  ``Symmetric preferences, asymmetric outcomes: Tipping dynamics in an open-city segregation model'' \\[0.5em]
  \large
  Fabio van Dissel, Tuan Minh Pham and Wout Merbis
\end{center}
\makeatother

\vspace{1em}


\section{Description of numerical algorithm}
\label{sec:appendix_implementation}

In this section we describe the stochastic simulation algorithm used to generate the numerical results presented in this work, in particular those underlying the finite-size scaling analysis. The dynamics are simulated using a standard Gillespie stochastic simulation algorithm \cite{gillespie1977exact} applied to a chemical reaction network defined on a two-dimensional square lattice of size $N=L\times L$.

Each lattice site $i$ can be in one of three states, $\sigma_i\in\{-1,0,1\}$, corresponding to an agent of type $B$, a vacant site, or an agent of type $A$, respectively. Interactions occur within the Moore neighborhood, such that each site has $n=8$ neighbors. At any instant, a site may undergo (i) neighborhood-independent birth--death reactions, or (ii) interaction-dependent in-group and out-group reactions involving its neighbors. The precise reaction channels and their associated rates are specified in the main text equations \eqref{death}-\eqref{split}.

For a given lattice configuration $\Sigma=\{\sigma_i\}$, all admissible reactions define a set of independent reaction channels indexed by $\alpha$, each characterized by a propensity (rate) $p_\alpha(\Sigma)$. The total propensity is given by
\begin{equation}
P(\Sigma)=\sum_{\alpha} p_\alpha(\Sigma),
\end{equation}
which determines both the waiting time to the next reaction event and the relative probability with which each reaction channel is selected. The resulting continuous-time Markov process is simulated using the Gillespie algorithm summarized in Algorithm~\ref{alg:gillespie}.

Initial conditions are generated by fixing an initial vacancy density $\rho_0$. Each lattice site is independently assigned to be vacant with probability $\rho_0$, and otherwise occupied by an $A$ or $B$ agent with equal probability. We verified that the choice of $\rho_0$ does not affect steady-state observables. Without loss of generality, the death rate is fixed to $\mu=1$, thereby setting the unit of time; all other rates are expressed relative to this value.

To ensure convergence to the steady state, each simulation is first evolved for $n_{\mathrm{therm}} = 180\,N$ reaction events. Observables are subsequently recorded every $18\,N$ reaction events. We verified that this sampling interval is sufficient to eliminate temporal correlations between successive measurements. Expectation values are obtained by averaging over many statistically independent steady-state configurations. For the finite-size scaling analysis, simulations were parallelized over 128 CPU cores, each providing an independent realizations evolving according to Algorithm~\ref{alg:gillespie}. All code created to run the simulation and analyze the results are published in an accompanying GitHub repository \cite{github_repo}.

\begin{algorithm}[H]
\caption{Gillespie algorithm for lattice reaction dynamics}
\label{alg:gillespie}
\begin{minipage}{.96\columnwidth}
\begin{algorithmic}[1]
\State Initialize lattice configuration $\Sigma$ and set $t=0$
\While{simulation not terminated}
    \State Enumerate all admissible reaction channels $\{\alpha\}$ and compute propensities $p_\alpha(\Sigma)$
    \State Compute total propensity $P=\sum_\alpha p_\alpha$
    \State Draw $r_1,r_2\sim\mathcal{U}(0,1)$
    \State Advance time: $t \leftarrow t + \frac{1}{P}\ln(1/r_1)$
    \State Select reaction $\alpha$ such that
    \[
    \sum_{\beta<\alpha} p_\beta < r_2 P \le \sum_{\beta\le\alpha} p_\beta
    \]
    \State Update configuration $\Sigma \leftarrow \Sigma  + \Delta \Sigma_\alpha$
    \State Record observables if required
\EndWhile
\end{algorithmic}
\end{minipage}
\end{algorithm}

\section{Comparing mean-field to stochastic solution}\label{sec:comparison}
To investigate the validity of the mean-field results we perform numerical simulations as outlined in section~\ref{sec:appendix_implementation}. We focus on two observables, to distinguish the different phases of the model. Namely, the net overdensity, explicitly defined on the lattice as
\begin{equation}
    \hat{M} =\frac{1}{N} \sum_i \sigma_i
    \label{eq:netmagnetization}
\end{equation}
Where the sum runs over all lattice sites and $\sigma_i = 1, -1, 0$ if lattice site $i$ is in the state $A_i$, $B_i$ or $0_i$ respectively. This observable should accurately distinguish between the Symmetric and Asymmetric phases and therefore serves as a useful order parameter. Next, we need an observable that accurately tracks the segregation on the lattice. Note that one can have segregation with no net magnetization. For this, we define something akin to the total internal energy of the lattice,
\begin{equation}
    \hat{E} = \frac{1}{8 N}\sum_{i,j} \sigma_i C^{ij}\sigma_j
    \label{eq:Energyobservable}
\end{equation}
Where $C^{ij}$ is the adjacency matrix. This should be an accurate proxy for segregation as it directly measures how many distinct agents are `living' next to each other.

\begin{figure}
    \centering
    \includegraphics[width = \textwidth]{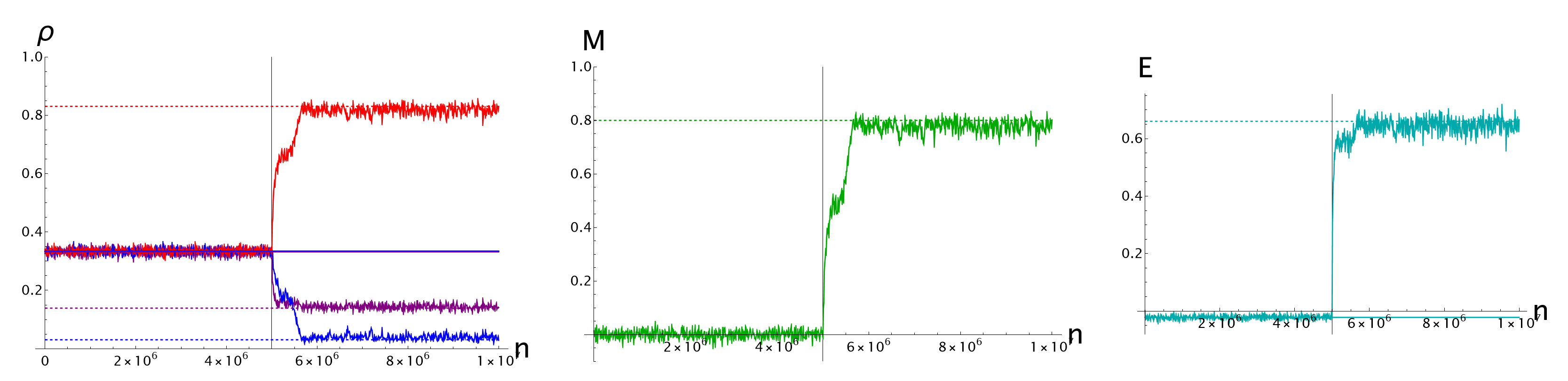}
    \caption{The average densities (left), net magnetization (middle) and the Ising energy (right) for the quench experiment described in the text. The mean-field expectation (2nd order) in the symmetric (full) and asymmetric (dashed) phases are also shown. The grid-line correspond to the step at which we change the value of $r_c$.}
    \label{fig:quench}
\end{figure}

We perform simulations of the full stochastic model. To investigate the transition between the two phases we initially set $r_c = \frac{r_s}{2}$, with $r_s = 0.2$. Then, after we have simulated $\mathcal{N}_{\rm sim} = 5 \cdot 10^6$ distinct reactions on our lattice we change the value of $r_c$ to $r_c = 2.5 >r_c^*$. We do this for a $N = 50 \times 50$ size lattice, with homogeneous initial conditions. In Fig.~\ref{fig:quench} we show the results of the average density of the different species in the lattice, and the average magnetization as defined in Eq.~\eqref{eq:netmagnetization}. It is clear that, before the quench, the system is in the symmetric phase (no net magnetization). After the quench the species $A$ starts to dominate and the system becomes ordered, agreeing with the (second-order) mean-field results.

In the right panel of Fig.~\ref{fig:quench} we show the energy as defined in Eq.~\eqref{eq:Energyobservable} for this same experiment. We see that initially the energy is slightly negative when $r_c < r_s$. This signals a different type of phase, which we refer to as anti-segregating. In this phase agents are more likely to live next to distinct neighbors, signaling a situation in which a neighborhood is very homogeneously mixed. This behavior is only visible in the second-order mean-field approach (see section \ref{appendix:Fockspace}). Note that the transition from negative to positive energy at $r_c = r_s$ is not critical. Nonetheless, we argue that this cross-over deserves some attention because of the socioeconomic context of this model.

\section{Fock space formulation of the model}
\label{appendix:Fockspace}
In this section we will work in the Fock space formulation of the CRN, as described in, for instance \cite{doi1976second,tauber2005applications,del2026field}. This allows us to derive the first and second order mean-field equations mentioned in the main text explicitly.
We start with defining the master equation for the CRN. The state of the system is a vector $\ket{\rho(t)}$, which is expanded as:
\begin{equation}\label{rhogen}
    \ket{\rho(t)} = \sum_{\Sigma} P(\Sigma,t) \prod_{i} (\a_i^\dagger)^{n_i^A}(\b_i^\dagger)^{n_i^B}(\v_i^\dagger)^{n_i^0}\ket{0}
\end{equation}
Where $P(\Sigma,t)$ gives the probability of observing the microscopic configuration $\Sigma = \{\sigma_i \}$ of lattice sites, each of which may be in any of the three states $A, B, 0$. We denote $\a_i, \a_i^\dagger$, $\b_i, \b_i^\dagger$ and $\v_i, \v_i^\dagger$ as the annihilation (creation) operators for $A_i, B_i$ and $0_i$ particles, respectively. They satisfy commutation relations:
\begin{equation}\label{commutators}
    [\a_i, \a_j^\dagger] = \delta_{ij} = [\b_i, \b_j^\dagger] = [\v_i, \v_j^\dagger]\,.
\end{equation}
The master equation is given as:
\begin{equation}\label{msteqn}
    \partial_t \ket{\rho(t)} = \hat{H} \ket{\rho(t)}\,,
\end{equation}
where the infinitesimal generator $\hat{H}$ is split into three parts:
\begin{equation}\label{generator}
    \hat{H} = \hat{H}_0 (T) + \hat{H}_c(T) + \hat{H}_s(T) \,.
\end{equation}
Each term represents a set of reactions, the first corresponds to the birth/death process at a given level of $T$ (we again set $\lambda =1$ by rescaling time):
\begin{align}\label{H0}
    \hat{H}_0 & = \sum_i \Big\{ T (\v_i^\dagger - \a_i^\dagger) \a_i + T (\v_i^\dagger - \b_i^\dagger) \b_i + (\a_i^\dagger +\b_i^\dagger - 2 \v_i^\dagger)\v_i \Big\}
\end{align}
The generator $\hat{H}_c$ represents the cloning reactions:
\begin{align}\label{Hcopy}
    \hat{H}_c & = r_c \sum_{\langle i,j\rangle} \Big\{ T (\v_i^\dagger\a_j^\dagger - \b_i^\dagger \a_j^\dagger) \b_i \a_j + T(\v_i^\dagger \b_j^\dagger - \a_i^\dagger \b_j^\dagger) \a_i \b_j + (\a_i^\dagger \a_j^\dagger - \v_i^\dagger \a_j^\dagger) \v_i \a_j + (\b_i^\dagger \b_j^\dagger - \v_i^\dagger \b_j^\dagger)\v_i \b_j  \Big\}
\end{align}
Finally, the generator $\hat{H}_s$ represents the splitting reactions:
\begin{align}\label{Hsplit}
    \hat{H}_s & = r_s \sum_{\langle i,j\rangle} \Big\{ T(\vidag \b_j^\dagger - \bidag \b_j^\dagger)\bi \b_j + T(\vidag \a_j^\dagger - \aidag \a_j^\dagger)\ai \a_j 
    + (\bidag \a_j^\dagger - \vidag \a_j^\dagger)\vi \a_j + (\aidag \b_j^\dagger - \vidag \b_j^\dagger)\vi \b_j \Big\}
\end{align}
Here $\langle i,j\rangle$ denotes nearest neighbors (within the Moore neighborhoods).

\subsection{Mean-field equations} \label{app:mean_field}
From the master equation \eqref{msteqn}, one may derive equations for the expectation value for node $i$ to be in the state $A$, denoted as $\rho_i^A(t) = \bra{1}\aidag\ai \ket{\rho(t)}$. The equation is easily derived using that $\partial_t \rho_i^A = \bra{1} \ai \hat{H}\ket{\rho(t)}$, together with the commutation relations \eqref{commutators}:
\begin{subequations}
    \label{FOequations}
\begin{align}
    \frac{d}{dt} \rho_i^A & = \rho_i^0 - T \rho_i^A + r_c \sum_{j \in \partial i} \left( \vev{\vi \hat{a}_j}- T\vev{\ai \hat{b}_j}\right) 
    + r_s \sum_{j \in \partial i} \left( \vev{\vi \hat{b}_j}- T\vev{\ai \hat{a}_j}\right)
\end{align}
Here $\partial i$ is the local neighborhood of node $i$ and $\vev{\vi \hat{a}_j}$ denotes the probability that nodes $i$ and $j$ form a $(0A)$-pair (and likewise for the other two point correlators). Using similar computations for $\rho^B_i$ and $\rho^0_i$ gives:
\begin{align}
    \frac{d}{dt} \rho_i^B & = \rho_i^0 - T \rho_i^B + r_c \sum_{j \in \partial i} \left( \vev{\vi \hat{b}_j}- T\vev{\bi \hat{a}_j}\right) 
    + r_s \sum_{j \in \partial i} \left( \vev{\vi \hat{a}_j}- T\vev{\bi \hat{b}_j}\right)\,, \\
    \frac{d}{dt} \rho_i^0 & = T (\rho_i^A + \rho_i^B) - 2 \rho_i^0 
    + r_c \sum_{j \in \partial i} \left(T \vev{\bi \hat{a}_j} + T \vev{\ai \hat{b}_j}- \vev{\vi \hat{a}_j} - \vev{\vi \hat{b}_j}\right) \\
    & \quad + r_s \sum_{j \in \partial i} \left(T \vev{\bi \hat{b}_j} + T \vev{\ai \hat{a}_j}- \vev{\vi \hat{a}_j} - \vev{\vi \hat{b}_j}\right) \nonumber\,.
\end{align}
\end{subequations}
These equations are exact, but do not form a closed system. In order to arrive at a single set of closed ODE's we make the mean-field assumption that each site is statistically independent and identically distributed. If this were true, we may write any pair expectation values as the product of the single nodes expectation values, i.e. $\vev{\vi \hat{a}_j} = \rho_i^0 \rho_j^A$. Furthermore, we can invoke translational invariance by supposing that each lattice site is identical, such that the labels $i,j$ become irrelevant. In that case, we arrive at the mean field equations for the probability of any site to be in the state $A,B$ or $0$:
\begin{subequations}
\begin{align}
    \frac{d}{dt} \rho^A & = \rho^0 - T \rho^A + 8 r_c \left(\rho^0 \rho^A - T \rho^A \rho^B\right) 
    + 8 r_s \left( \rho^0 \rho^B - T\rho^A \rho^A \right)  \\
    \frac{d}{dt} \rho^B & = \rho^0 - T \rho^B + 8 r_c \left( \rho^0 \rho^B - T\rho^A \rho^B\right)
    + 8 r_s \left( \rho^0 \rho^A - T \rho^B \rho^B \right)\,, \\
    \frac{d}{dt} \rho^0 & = T (\rho^A + \rho^B) - 2 \rho^0 
    + 8 r_c \left(2 T \rho^A \rho^B  - \rho^0 (\rho^A + \rho^B) \right)  \\
    & \quad + 8 r_s \left(T (\rho^A \rho^A+ \rho^B \rho^B)- \rho^0 (\rho^A + \rho^B) \right) \nonumber\,.
\end{align}
\end{subequations}
These formula's are equivalent to equations \eqref{meanfieldequations}.

\subsection{Second-order mean field equations}\label{sec:2ndorderMF}
In the previous subsection, we closed the equations at the first order, but a better mean-field approximation can be obtained when closing the equations at the second order \cite{kuehn2024preserving,Wuyts2022}. To this end, we first formulate the equations for the second moments (or the two-point correlation functions), which represent the joint probabilities of forming a pair of specific agents. These equations are expressed in terms of pairs and triplets on the lattice.
\begin{subequations} \label{twopointfunctions}
\begin{align}
    & \frac{d}{dt} \vev{\ai \a_j} = (1+r_c) (\vev{\ai \v_j} + \vev{\vi \a_j}) - 2T(1+r_s) \vev{ \ai \a_j} + r_c \big[\sum_{k \in \partial i}\{ \vev{\vi \a_k \a_j } - T \vev{\ai \b_k \a_j}\} + \sum_{k \in \partial j}\{ \vev{\ai \a_k \v_j } - T \vev{\ai \b_k \a_j} \} \big] \nonumber \\
    & \quad + r_s \big[\sum_{k \in \partial i}\{ \vev{\vi \b_k \a_j } - T \vev{\ai \a_k \a_j} \} + \sum_{k \in \partial j}\{ \vev{\ai \b_k \v_j } - T \vev{\ai \a_k \a_j} \} \big] \,,  \\
    & \frac{d}{dt} \vev{\ai \v_j} =  \vev{\vi \v_j} + T(1 + r_c)\vev{\ai \b_j} + T(1 + r_s) \vev{\ai \a_j}  - (T+2+r_c+r_s) \vev{ \ai \v_j} \nonumber \\
    & \quad + \sum_{k \in \partial i}\{ r_c \vev{\vi \a_k \v_j } + r_s \vev{\vi \b_k \v_j } - T r_c \vev{\ai \b_k \v_j} - T r_s \vev{\ai \a_k \v_j }\} \nonumber \\
    & \quad + \sum_{k \in \partial j} \Big\{ T \big[ r_c \vev{\ai \b_k \a_j } + r_c \vev{\ai \a_k \b_j} + r_s \vev{\ai \a_k \a_j } + r_s \vev{\ai \b_k \b_j } \big] - (r_s + r_c) [\vev{\ai \a_k \v_j } +  \vev{\ai \b_k \v_j } ] \Big\}\,,  \\
    & \frac{d}{dt} \vev{\vi \v_j} =  T \big[\vev{\ai \v_j} + \vev{\bi \v_j} + \vev{\vi \a_j} + \vev{\vi \b_j} \big] - 4 \vev{\vi \v_j} \nonumber \\
    & \quad + \sum_{k \in \partial i} \Big\{ T \big[ r_c \vev{\ai \b_k \v_j } + r_c \vev{\bi \a_k \v_j } + r_s \vev{\ai \a_k \v_j } + r_s \vev{\bi \b_k \v_j } \big] - (r_c +r_s) \big[  \vev{\vi \a_k \v_j} + \vev{\vi \b_k \v_j } \big] \Big\} \nonumber \\
    & \quad + \sum_{k \in \partial j} \Big\{ T \big[ r_c \vev{\vi \b_k \a_j } + r_c \vev{\vi \a_k \b_j } + r_s \vev{\vi \a_k \a_j } + r_s \vev{\vi \b_k \b_j } \big]  - (r_c +r_s) \big[  \vev{\vi \a_k \v_j} + \vev{\vi \b_k \v_j } \big] \Big\}\,, \\
    & \frac{d}{dt} \vev{\ai \b_j} = (1+r_s) (\vev{\ai \v_j} + \vev{\vi \b_j}) - 2T(1+r_c) \vev{ \ai \b_j} 
    + r_c \big[\sum_{k \in \partial i}\{ \vev{\vi \a_k \b_j } - T \vev{\ai \b_k \b_j}\} + \sum_{k \in \partial j}\{ \vev{\ai \b_k \v_j } - T \vev{\ai \a_k \b_j} \} \big] \nonumber \\
    & \quad + r_s \big[\sum_{k \in \partial i}\{ \vev{\vi \b_k \b_j } - T \vev{\ai \a_k \b_j} \} + \sum_{k \in \partial j}\{ \vev{\ai \a_k \v_j } - T \vev{\ai \b_k \b_j} \} \big] \,.
\end{align}    
\end{subequations}
The equations for $\vev{\b_i \b_j}, \vev{\bi\v_j}$ may be obtained from these expressions by replacing $\a \leftrightarrow \b$. Likewise, expressions for $\vev{\vi \a_j}, \vev{\vi \b_j}$ and $\vev{\b_i \a_j}$ are obtained by swapping lattice labels $i \leftrightarrow j $. In addition to these equations, the finite volume constraint imposes that the sum of all possible two-point correlation functions is one, hence we may always choose to express one two-point correlation function in terms of the others.
Together with equations \eqref{FOequations}, these equations provide the exact description of the one- and two-point correlation functions of the system. However, to form a closed set of equations, one would have to approximate the three-point correlation functions in terms of the one and two-point correlators. 

Before closing the set of equations, we note that owing to our choice of Moore neighborhoods, there are two statistically distinct geometries possible for the pair correlation functions: the pair could form a straight edge (if they are left, right, up or down neighbors), or the pair could form a diagonal edge (if they are neighbors connecting to the 4 corners of the Moore neighborhood). These choices are distinct, as straight edges are part of four closed triangles, while diagonal edges are part of only 2 closed triangles. So for each possible pair of nodes, we separate the correlation functions into $\vev{}_{\rm s}$ if they form up-, down-, left- or right neighbors and $\vev{}_{\rm d}$ if they form diagonal neighbors.

When approximating the three-point functions to close the set of equations, we distinguish between two further scenarios. First, the triplet $i,j,k$ may form an open triangle, such that one node is connected to the two others, but its neighbors are not connected to each other. The second case is when the triplet $i,j,k$ forms a closed triangle and all nodes are connected to each other. The rules for closing the moment equations in these two cases are detailed as the examples (1) and (2) in \cite{Wuyts2022}. Specifically, for an open triangle with $j$ being central node, we approximate the third moment as 
\begin{equation}\label{eq:open}
    \vev{\hat{x}_i \hat{y}_j \hat{z}_k} \sim \frac{\vev{\hat{x}_i \hat{y}_j} \vev{\hat{y}_j \hat{z}_k}}{\vev{\hat{y}_j}}\,.
\end{equation} 
When instead nodes $i,j,k$ form a closed triangle (such that all nodes are connected to each other), we approximate the correlation function as:
\begin{equation}\label{eq:closed}
    \vev{\hat{x}_i \hat{y}_j \hat{z}_k} \sim \frac{\vev{\hat{x}_i \hat{y}_j} \vev{\hat{y}_j \hat{z}_k}\vev{\hat{z}_k \hat{x}_i}}{\vev{\hat{x}_i} \vev{\hat{y}_j} \vev{\hat{z}_k}}\,.
\end{equation}
which is a common closure scheme in epidemic models \cite{Keeling}.

\begin{figure}
    \centering
    \includegraphics[width=0.8\linewidth]{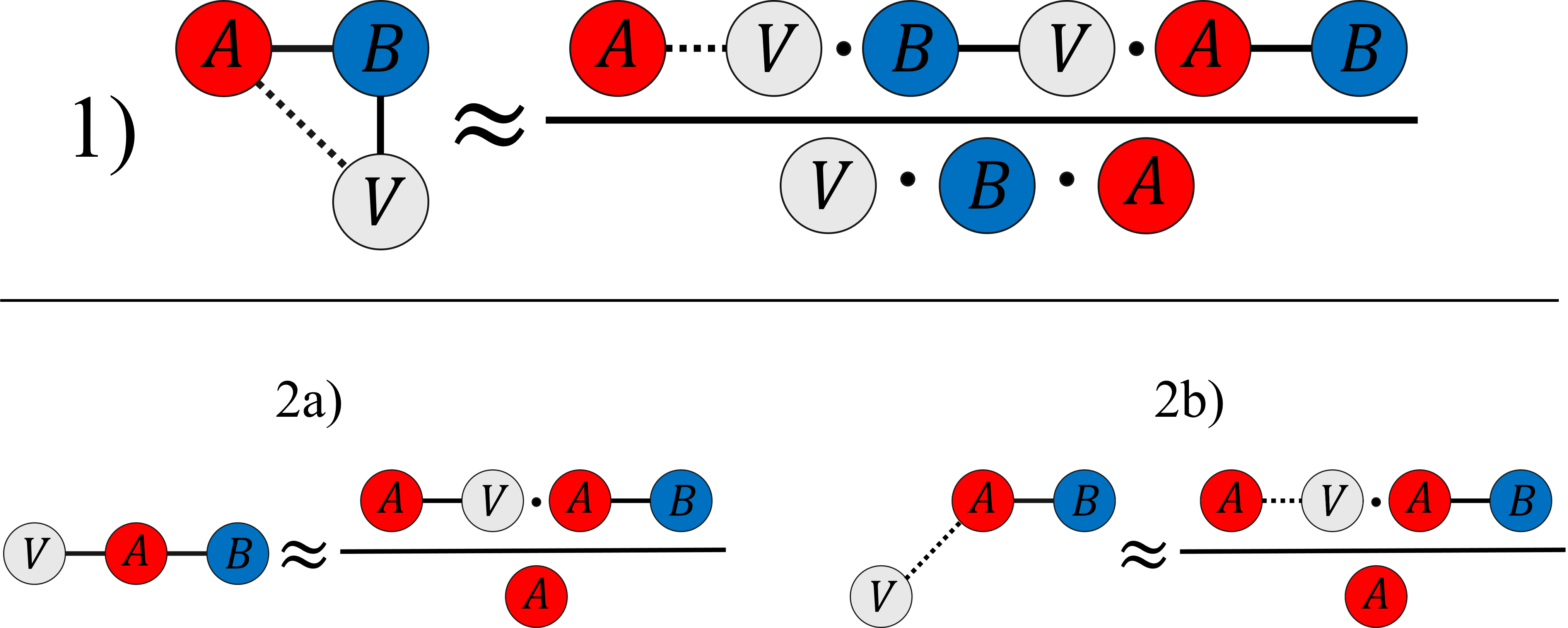}
    \caption{A sketch illustrating how we perform moment closure of three-point correlation functions. \textit{1:} If the sites in the three-point function form a closed triangle, it consists of two straight edges (full black line) and one diagonal edge (dashed black line). The expectation value is approximated as in \eqref{eq:closed}. \textit{2:} If on the other hand the three-point function is open we close as in \eqref{eq:open}. An open triangle can consist of two straight edges (\textit{2a}), one diagonal and one straight edge (\textit{2b}) or two diagonal edges (not shown).}
    \label{fig:momentclosure}
\end{figure}

\begin{figure}
    \centering
    \includegraphics[width=0.8\linewidth]{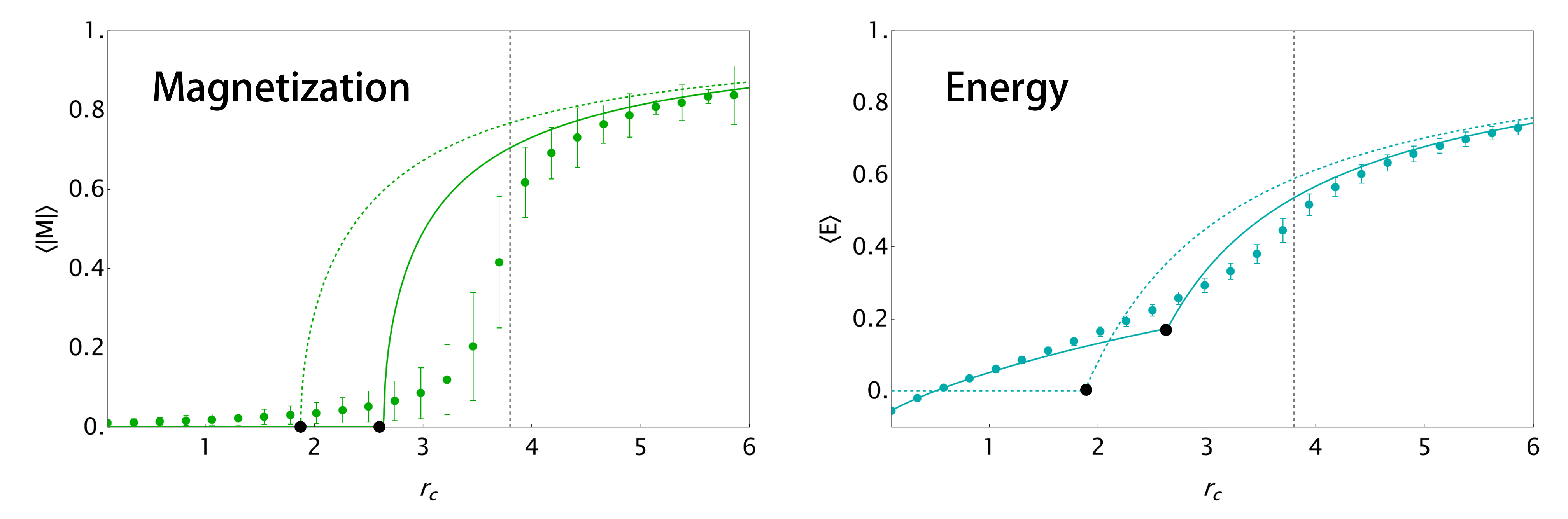}
    \caption{Comparison in expectation value for the absolute magnetization per site (left) and the energy per site (right) between the moment closure of the mean field equations at first (dashed line) and second (full line) order. Plots shown here are for $r_s = 0.5$ and $T = 1$. The scatter data corresponds to results from the stochastic model, averaged over $10^4$ independent grids with grid-size $N = 50 \times 50$. We indicate the critical value of $r_c^*$ with the dashed gridline.} 
    \label{fig:secondorderMF2}
\end{figure}

Combining the two distinct geometries for the two-point functions (straight or diagonal) with the two closure schemes for three-point functions (open or close triangles) leads to the following replacement rules for the sum over neighbors $k$ of any node pair $i$ and $j$ (see also Fig.~\ref{fig:momentclosure}). Whenever the nodes $i$ and $j$ are straight neighbors, the node $k \in \partial i$ can form a closed triangle with $i$ and $j$ in four ways and an open triangle in three ways. Keeping track of the straight and diagonal connections leads to:
\begin{equation}\label{triangle_replace_straigth}
    \sum_{k \in \partial i} \vev{\hat{x}_i \hat{y}_j \hat{z}_k} = 2 \frac{\vev{\hat{x} \hat{y}}_{\rm s} \vev{\hat{x} \hat{z}}_{\rm d}}{\vev{\hat{x}}} + \frac{\vev{\hat{x} \hat{y}}_{\rm s} \vev{\hat{x} \hat{z}}_{\rm s}}{\vev{\hat{x}}} + 2 \frac{\vev{\hat{x} \hat{y}}_{\rm s} \big(\vev{\hat{y}\hat{z}}_{\rm s}\vev{\hat{z} \hat{x}}_{\rm d} + \vev{\hat{y}\hat{z}}_{\rm d}\vev{\hat{z} \hat{x}}_{\rm s}\big)}{\vev{\hat{x}} \vev{\hat{y}} \vev{\hat{z}}}  
\end{equation}
However, if the node pair $i$ and $j$ are diagonal neighbors, the node $k \in \partial i$ forms a closed triangle with $i$ and $j$ in only 2 ways, and creates an open triangle in 5 ways. The replacement rule is then:
\begin{equation}\label{triangle_replace_diagonal}
    \sum_{k \in \partial i} \vev{\hat{x}_i \hat{y}_j \hat{z}_k} = 2 \frac{\vev{\hat{x} \hat{y}}_{\rm d} \vev{\hat{x} \hat{z}}_{\rm s}}{\vev{\hat{x}}} + 3 \frac{\vev{\hat{x} \hat{y}}_{\rm d} \vev{\hat{x} \hat{z}}_{\rm d}}{\vev{\hat{x}}} + 2 \frac{\vev{\hat{x} \hat{y}}_{\rm d} \vev{\hat{y}\hat{z}}_{\rm s}\vev{\hat{z} \hat{x}}_{\rm s} }{\vev{\hat{x}} \vev{\hat{y}} \vev{\hat{z}}}  
\end{equation}
Substituting these replacements rules in equations \eqref{twopointfunctions} and separating out the straight and diagonal edges is straight-forward albeit tedious. It results in a set of 15 equations; three for the one point functions $\vev{\a}, \vev{\b}, \vev{\v}$, six for all possible straight edges $\vev{\a\a}_{\rm s}, \vev{\a\b}_{\rm s}, \vev{\a\v}_{\rm s}, \vev{\b\v}_{\rm s}, \vev{\b\b}_{\rm s}, \vev{\v\v}_{\rm s}$ and six for the diagonal edges. Among these equations, three are solved by imposing the volume exclusion constraints $\vev{\v} = 1-\vev{\a}-\vev{\b}, \vev{\v \v}_{\rm s} = 1 - \vev{\a\a}_{\rm s} - \vev{\b\b}_{\rm s} - 2\vev{\a\b}_{\rm s} - 2\vev{\a\v}_{\rm s} - 2\vev{\b\v}_{\rm s}$, and likewise for the diagonal edges. 
The resulting set of equations are solved by numerical integration, using the first order mean-field solution as initial conditions. This allows us to obtain better estimates of the absolute magnetization and the energy, as shown in Figure \ref{fig:secondorderMF2}. The code for numerical integration of these equations is included in the GitHub repository \cite{github_repo}.

\section{Finite-size robustness analysis}
\label{app:finite-size-robustness}

To assess the sensitivity of the finite-size scaling analysis to the range of lattice sizes included in the fits, we performed additional robustness checks for the cases \(r_s=0.1\) and \(r_s=0\). In the main analysis, the critical exponents were extracted from simulations with linear system sizes \(L=20,25,30,35,40,45\). We extended these data sets by including additional simulations at \(L=50,55,60\), and repeated the full finite-size scaling procedure.

For each fitting window, the critical point \(r_c^\ast\) was re-estimated from the Binder cumulant crossing using the same variance-minimization procedure as in the main text. The observables were then interpolated to the corresponding window-specific value of \(r_c^\ast\), and the exponent ratios were obtained from log-log fits at criticality. The correlation-length exponent \(\nu\) was determined independently by optimizing the collapse of the magnetization curves. Note that values reported here are obtained from the analysis on the simulated data, without bootstrapping for estimating uncertainties. This may result in deviations from the values reported in Table \ref{tab:criticalexponents} in the main text.

\subsection*{Robustness at \(r_s=0.1\)}

For \(r_s=0.1\), the comparison between the original window and the full extended window is summarized as follows. Using the original range \(L=20\text{--}45\), we find
\begin{equation}
r_c^\ast = 1.38695,\quad
\beta/\nu = -0.13007,\quad
\gamma/\nu = 1.73955,\quad
\alpha/\nu = 0.26463,\quad
\epsilon/\nu = -1.83256, \quad
\nu = 1.03155 .
\end{equation}
Including the additional larger system sizes up to \(L=60\) gives
\begin{equation}
r_c^\ast = 1.38988,\quad
\beta/\nu = -0.12824,\quad
\gamma/\nu = 1.74453,\quad
\alpha/\nu = 0.26287,\quad
\epsilon/\nu = -1.82892,\quad
\nu = 1.060547 .
\end{equation}
The corresponding relative changes are small: approximately \(1.4\%\) for \(\beta/\nu\), \(0.3\%\) for \(\gamma/\nu\), \(0.7\%\) for \(\alpha/\nu\), and \(2.8\%\) for \(\nu\). Thus, extending the upper end of the system-size range does not significantly alter the exponent estimates. In particular, the enlarged data set does not show a systematic drift of the exponents toward the two-dimensional Ising values.

A more detailed overview is given in Table~\ref{tab:finite_size_robustness_rs01}. We report both an upper-cutoff scan, in which the smallest system size is kept fixed while progressively increasing \(L_{\max}\), and a lower-cutoff scan, in which the largest available system size is kept fixed while progressively increasing \(L_{\min}\).

\begin{table}[t]
\centering
\begin{tabular}{c c c c c c c}
\hline\hline
Window type & \(L\)-window & \(r_c^\ast\) & \(\beta/\nu\) & \(\gamma/\nu\) & \(\alpha/\nu\) &\(\nu\) \\
\hline
Original & \(20\text{--}45\) & 1.386946 & -0.130075 & 1.739547 & 0.264627 &  1.031553 \\
Full & \(20\text{--}60\) & 1.389884 & -0.128239 & 1.744529 & 0.262874 &  1.060547 \\
\hline
Upper cutoff & \(20\text{--}35\) & 1.386026 & -0.129605 & 1.742797 & 0.275581 & 1.048964 \\
Upper cutoff & \(20\text{--}40\) & 1.384625 & -0.132275 & 1.735856 & 0.270319 & 1.022459 \\
Upper cutoff & \(20\text{--}50\) & 1.387805 & -0.129058 & 1.742665 & 0.263814 & 1.042221 \\
Upper cutoff & \(20\text{--}55\) & 1.388086 & -0.131328 & 1.739259 & 0.262499 & 1.044124 \\
\hline
Lower cutoff & \(25\text{--}60\) & 1.391114 & -0.128173 & 1.745158 & 0.258669 & 1.069071 \\
Lower cutoff & \(30\text{--}60\) & 1.392337 & -0.127948 & 1.745961 & 0.259500 & 1.081009 \\
Lower cutoff & \(35\text{--}60\) & 1.404472 & -0.096191 & 1.788940 & 0.265058 & 1.191127 \\
Lower cutoff & \(40\text{--}60\) & 1.406022 & -0.095630 & 1.789244 & 0.266174 & 1.227072 \\
Lower cutoff & \(45\text{--}60\) & 1.407224 & -0.093081 & 1.795524 & 0.273589 & 1.243050 \\
\hline\hline
\end{tabular}
\caption{
Finite-size robustness of the exponent estimates for \(r_s=0.1\). The upper-cutoff scan keeps the lower end of the fitting range fixed and progressively includes larger system sizes. The lower-cutoff scan keeps the largest available size fixed and progressively removes smaller system sizes. Each row uses a window-specific estimate of \(r_c^\ast\).
}
\label{tab:finite_size_robustness_rs01}
\end{table}

The upper-cutoff scan shows that the exponent ratios are stable when larger system sizes are added. Across these fits, \(\beta/\nu\) remains close to \(0.13\), \(\gamma/\nu\) remains close to \(1.74\), and \(\alpha/\nu\) remains positive and close to \(0.26\text{--}0.28\). This indicates that the previously reported exponents are not an artifact of having stopped the analysis at \(L=45\).

The lower-cutoff scan shows a stronger drift once the smallest sizes are removed, especially for \(L_{\min}\geq 35\). However, these restricted fits are based on a much narrower range of system sizes. For example, the window \(L=45\text{--}60\) contains only four data points and has a small logarithmic lever arm for determining power-law slopes. In addition, the Binder estimate of \(r_c^\ast\) shifts noticeably in these restricted windows, from \(r_c^\ast\simeq 1.392\) for \(L_{\min}=30\) to \(r_c^\ast\simeq 1.407\) for \(L_{\min}=45\). Since the observables are evaluated at the window-specific critical point, this drift in \(r_c^\ast\) feeds directly into the fitted exponent ratios.

\subsection*{Robustness at \(r_s=0\)}

We performed the same analysis for \(r_s=0\), which is the case shown explicitly in the Binder-cumulant crossing and scaling-collapse figures in the main text. Using the original range \(L=20\text{--}45\), we find
\begin{equation}
r_c^\ast = 0.78739,\quad
\beta/\nu = -0.13013,\quad
\gamma/\nu = 1.73474,\quad
\alpha/\nu = 0.27161,\quad
\epsilon/\nu = -1.82618,\quad
\nu = 1.09689 .
\end{equation}
Including the additional larger system sizes up to \(L=60\) gives
\begin{equation}
r_c^\ast = 0.78948,\quad
\beta/\nu = -0.12549,\quad
\gamma/\nu = 1.74913,\quad
\alpha/\nu = 0.27014,\quad
\epsilon/\nu = -1.83038,\quad
\nu = 1.10283 .
\end{equation}
The corresponding changes are again modest: \(r_c^\ast\) shifts by approximately \(0.27\%\), while the exponent ratios change by approximately \(3.6\%\) for \(\beta/\nu\), \(0.8\%\) for \(\gamma/\nu\), \(0.5\%\) for \(\alpha/\nu\), and \(0.5\%\) for \(\nu\). The vacancy-fluctuation exponent ratio is similarly stable. Thus, the inclusion of \(L=50,55,60\) does not qualitatively change the finite-size scaling estimates for \(r_s=0\).
A detailed overview of the window dependence for \(r_s=0\) is given in Table~\ref{tab:finite_size_robustness_rs00}.

\begin{table}[t]
\centering
\begin{tabular}{c c c c c c c}
\hline\hline
Window type & \(L\)-window & \(r_c^\ast\) & \(\beta/\nu\) & \(\gamma/\nu\) & \(\alpha/\nu\) & \(\nu\) \\
\hline
Original & \(20\text{--}45\) & 0.787393 & -0.130129 & 1.734736 & 0.271607 & 1.096894 \\
Full & \(20\text{--}60\) & 0.789478 & -0.125486 & 1.749127 & 0.270144 & 1.102832 \\
\hline
Upper cutoff & \(20\text{--}35\) & 0.783618 & -0.136797 & 1.721781 & 0.278827 & 1.075518 \\
Upper cutoff & \(20\text{--}40\) & 0.785211 & -0.134085 & 1.725577 & 0.275139 & 1.086573 \\
Upper cutoff & \(20\text{--}50\) & 0.788744 & -0.126559 & 1.744422 & 0.273748 & 1.103205 \\
Upper cutoff & \(20\text{--}55\) & 0.788747 & -0.126691 & 1.746169 & 0.272193 & 1.092804 \\
\hline
Lower cutoff & \(25\text{--}60\) & 0.790744 & -0.123081 & 1.754983 & 0.267911 & 1.102954 \\
Lower cutoff & \(30\text{--}60\) & 0.791287 & -0.121284 & 1.757778 & 0.264650 & 1.105099 \\
Lower cutoff & \(35\text{--}60\) & 0.792204 & -0.117930 & 1.763339 & 0.266301 & 1.108719 \\
Lower cutoff & \(40\text{--}60\) & 0.793413 & -0.112685 & 1.768435 & 0.267661 & 1.113491 \\
Lower cutoff & \(45\text{--}60\) & 0.792703 & -0.110769 & 1.766294 & 0.263455 & 1.142352 \\
\hline\hline
\end{tabular}
\caption{
Finite-size robustness of the exponent estimates for \(r_s=0\). The upper-cutoff scan keeps the lower end of the fitting range fixed and progressively includes larger system sizes. The lower-cutoff scan keeps the largest available size fixed and progressively removes smaller system sizes. Each row uses a window-specific estimate of \(r_c^\ast\).
}
\label{tab:finite_size_robustness_rs00}
\end{table}

The behavior at \(r_s=0\) is consistent with the conclusions drawn from \(r_s=0.1\). In the upper-cutoff scan, adding larger system sizes leads only to a weak drift of the exponent estimates. The estimate of \(\beta/\nu\) decreases slightly when the full range up to \(L=60\) is included, while \(\gamma/\nu\) increases slightly; however, both remain close to the values obtained from the original fitting window. The heat-capacity exponent ratio remains positive and close to \(\alpha/\nu\simeq 0.27\), again showing no indication of crossing over to a logarithmic divergence over the accessible size range.

The lower-cutoff scan for \(r_s=0\) displays a smoother drift than the corresponding scan at \(r_s=0.1\). As the smallest system sizes are removed, \(\beta/\nu\) decreases from \(0.123\) at \(L_{\min}=25\) to \(0.111\) at \(L_{\min}=45\), while \(\gamma/\nu\) increases from \(1.755\) to \(1.766\). The estimate of \(\nu\) also increases mildly, reaching \(\nu\simeq 1.14\) for the narrowest window. As before, these restricted windows have a reduced logarithmic lever arm and should therefore be interpreted primarily as a diagnostic of residual fitting-window dependence.

Taken together, the \(r_s=0\) and \(r_s=0.1\) robustness checks support the same conclusion. The most relevant comparison is the extension of the original window \(L=20\text{--}45\) to the full window \(L=20\text{--}60\), for which the exponent estimates change only weakly. Within the enlarged system-size range, we find no evidence that the measured critical behavior drifts toward the standard two-dimensional Ising universality class. In particular, the heat-capacity scaling remains compatible with a positive power-law exponent rather than the logarithmic behavior expected for the two-dimensional Ising model.

\end{document}